\documentclass[aps,prd,preprintnumbers,showpacs,showkeys,nofootinbib,superscriptaddress,fleqn,floatfix,tightenlines,10pt]{revtex4-2} 
\usepackage{amsmath,amsfonts,amssymb,amscd,amsxtra,amsthm}
\usepackage{dsfont}
\usepackage{graphicx}  
\usepackage{epstopdf}
\usepackage{dcolumn}  
\usepackage{bm}          
\usepackage{slashed}
\usepackage[utf8]{inputenc}
\usepackage{booktabs}
\usepackage[normalem]{ulem} 
\usepackage[dvipsnames]{xcolor} 
\usepackage{enumitem}
\usepackage{array}
\usepackage{slashed}
\usepackage{tikz}
\usepackage{float}
\usepackage{multirow}
\usepackage{booktabs} 
\usepackage{slashed}

\renewcommand\sout{\bgroup \color{red} \ULdepth=-.5ex \ULset}

\begin{document}
\preprint{INHA-NTG-05/2024}
\title{\Large Production mechanism of the hidden charm pentaquark
  states $P_{c\bar{c}}$}  

\author{Samson Clymton}
\email[E-mail: ]{sclymton@inha.edu}
\affiliation{Department of Physics, Inha University,
Incheon 22212, Republic of Korea }

\author{Hyun-Chul Kim}
\email[E-mail: ]{hchkim@inha.ac.kr}
\affiliation{Department of Physics, Inha University,
Incheon 22212, Republic of Korea }
\affiliation{School of Physics, Korea Institute for Advanced Study 
  (KIAS), Seoul 02455, Republic of Korea}

\author{Terry Mart}
\email[E-mail: ]{terry.mart@sci.ui.ac.id}
\affiliation{Departemen Fisika, FMIPA, Universitas Indonesia, Depok
  16424, Indonesia} 
\date{\today}
\begin{abstract}
We investigate hidden-charm pentaquark states using an off-shell
coupled-channel formalism involving heavy meson and singly heavy
baryon scattering. Our approach utilizes an effective Lagrangian to
construct the kernel amplitudes, which respect both heavy quark
symmetry and hidden local symmetry. After solving the coupled
integral equations, we obtain the transition amplitudes for $J/\psi
N$ scattering and various heavy meson and singly heavy baryon
scattering processes. We identify seven distinct peaks related to
molecular states of heavy mesons $\bar{D}$ ($\bar{D}^*$) and singly
heavy baryons $\Sigma_c$ ($\Sigma_c^*$). Four of these peaks can be
associated with the known $P_{c\bar{c}}$ states: $P_{c\bar{c}}(4312)$,
$P_{c\bar{c}}(4380)$, $P_{c\bar{c}}(4440)$, and
$P_{c\bar{c}}(4457)$. We predict two additional resonances with
masses around 4.5 GeV, which we interpret as $\overline{D}^*
\Sigma_c^*$ molecular states, and identify one cusp
structure. Additionally, we predict two $P$-wave pentaquark states
with positive parity, which may be candidates for genuine pentaquark
configurations. Notably, these pentaquark states 
undergo significant modifications in the $J/\psi N$ elastic
channel, with some even disappearing due to interference from the
positive parity channel. The present investigation may provide
insight into the absence of pentaquark states in $J/\psi$
photoproduction observed by the GlueX collaboration. 
\end{abstract}
\maketitle

\section{Introduction}
The discovery of the pentaquarks by the LHCb collaboration has
invigorated research in heavy exotic baryon spectroscopy. Pentaquarks
with the minimal quark content $uudc\bar{c}$ were first identified in
the $J/\psi p$ invariant mass spectrum from $\Lambda_b^0 \to J/\psi p
K^{-}$ decays~\cite{LHCb:2015yax,LHCb:2019kea}. To date, four states
have been observed: three narrow states below the $\bar{D}^*
\Sigma_c$ threshold and one broad state below the
$\bar{D}\Sigma_c^*$ threshold. Recently, a new state designated as
$P_{c\bar{c}}(4330)$ was identified in $B_s^0\to J/\psi p\bar{p}$ decays, while
the previously observed $P_{c\bar{c}}(4312)$ was notably
absent~\cite{LHCb:2021chn}~\footnote{The naming convention of heavy
  pentaquarks is not yet settled. While the LHCb Collaboration
  suggested a naming convention for exotic
  hadrons~\cite{Gershon:2022xnn}, the Particle Data Group (PDG) uses a
  different one~\cite{PDG}. In the current work, we will follow the
  PDG convention.}. 
Intriguingly, the GlueX collaboration has found no evidence for these
pentaquark states in recent $J/\psi$ photoproduction experiments on
protons~\cite{GlueX:2019mkq,GlueX:2023pev}. Rather than invalidating 
previous findings, these seemingly conflicting results offer an
opportunity to deepen our understanding of pentaquark nature. 
The LHCb Collaboration has also identified the strangeness partner of
the pentaquark in the $J/\psi\Lambda$ invariant mass spectrum from
$\Xi_b^-\to J/\psi\Lambda K^-$ decays~\cite{LHCb:2020jpq}. Recently, 
they discovered the strangeness partner of $P_{c\bar{c}}$(4330) at a nearly 
identical mass, just 5 MeV higher~\cite{LHCb:2022ogu}. Investigations
into pentaquark spectra with strangeness $-2$ and $-3$ are ongoing,
with the CMS collaboration recently observing the decay
$\Lambda_b^0\to J/\psi\Xi^-K^+$~\cite{CMS:2024vnm}. However, low yield 
and poor resolution precluded observation of a clear spectrum in the
$J/\psi\Xi^-$ invariant mass.

Since the LHCb discovery of hidden charm pentaquarks $P_{c\bar{c}}$, a
plethora of theoretical works has been proposed to explain their
nature. Two distinct approaches suggest that $P_{c\bar{c}}$'s are
molecular states, lying below the thresholds of various heavy mesons
and singly heavy baryons. The first model posits bound states arising
from quark potential models in configuration space~\cite{Liu:2019zvb,
  Yalikun:2021bfm}. The second interprets them as poles in the lower
Riemann sheet, generated by non-perturbatively produced scattering
matrices~\cite{He:2019ify, Wang:2022oof, Shen:2024nck}. 
Furthermore, constituent quark interpretations offer another potential
scheme for explaining the pentaquark spectrum~\cite{Maiani:2015vwa,
  Lebed:2015tna, Wang:2015epa}. Additionally, alternative hypotheses
suggest that the peak structure is a consequence of kinematic
singularities~\cite{Liu:2015fea, Guo:2015umn, Bayar:2016ftu,
  Nakamura:2021qvy,Nakamura:2021dix}. It is not possible to
distinguish these schemes from the LHCb data alone. However, results
from the GlueX collaboration allow some constituent pentaquark spectra
to be ruled out due to their absence in $J/\psi$ photoproduction. 
In contrast, the triangle singularity scheme offers a partial
explanation for the disappearance of $P_{c\bar{c}}$ states observed by
the GlueX collaboration. This is attributed to the inability of
photoproduction to generate the double triangle diagram, as pointed out
in Ref.~\cite{Nakamura:2021qvy}. On the other hand, in the context of
the molecular picture, no mechanism has yet been identified that can
explain this disappearance. 

In the current work, we investigate hidden-charm pentaquark states
using heavy meson and singly heavy baryon scattering in an off-shell
coupled channel formalism. To facilitate understanding of experimental
findings, we also include charmonium-nucleon scattering. The
transition amplitudes are generated by solving coupled integral equations,
with kernel amplitudes constructed from meson-exchange diagrams. These
processes are governed by an effective Lagrangian that respects both
heavy quark symmetry and hidden local symmetry. 
Our analysis yields seven distinct peaks are related to
molecular states of the heavy mesons $\bar{D}$ ($\bar{D}^*$) and
singly heavy baryons $\Sigma_c$ ($\Sigma_c^*$). Four of 
these peaks can be identified with the known $P_{c\bar{c}}$ states:
$P_{c\bar{c}}(4312)$, $P_{c\bar{c}}(4380)$, $P_{c\bar{c}}(4440)$, and
$P_{c\bar{c}}(4457)$. The other three resonances, with masses around
4.5 GeV, are predicted to be $\overline{D}^*\Sigma_c^*$ molecular
states and are yet to be discovered. 
Notably, these peaks undergo significant modifications in the $J/\psi
N$ elastic channel, with some even disappearing due to interference
from the positive parity channel. This phenomenon may provide insight
into the absence of pentaquark states in $J/\psi$ photoproduction,
where elastic $J/\psi N$ scattering plays a crucial role. A more
quantitative explanation requires comparison of our theoretical model
with experimental data. This will be the subject of future studies,
employing more rigorous fitting strategies to provide a more
comprehensive and valid explanation of the observed phenomena. 

The current work is organized as follows: In Section~\ref{sec:2}, we
present the off-shell coupled-channel formalism used to study the
hidden-charm pentaquark states. This includes the effective
Lagrangian, the partial-wave expansion of the scattering amplitude,
and the method for solving the coupled integral
equations. Section~\ref{sec:3} is devoted to the results and
discussions, where we analyze the scattering matrices for both
negative and positive parity states, identify the pole positions, and
compare our findings with experimental data. We also address the
apparent conflict between LHCb and GlueX results, proposing a
qualitative explanation based on our molecular scheme. Finally, we
summarize our findings and conclude in Section~\ref{sec:4}, discussing
the implications of our results and outlining future directions.  

\section{Coupled-channel formalism\label{sec:2}} 
\begin{figure}[htbp]
  \centering
  \includegraphics[scale=1.0]{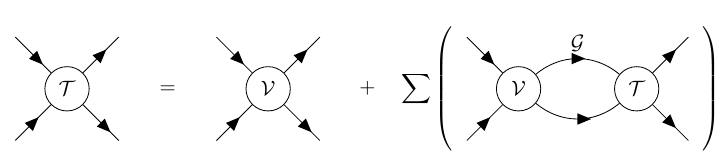}
  \caption{Graphical representation of the coupled integral
          scattering equation.}  
  \label{fig:1}
\end{figure}

The scattering amplitude is defined as 
\begin{align}
\mathcal{S}_{fi} = \delta_{fi} - i (2\pi)^4 \delta(P_f - P_i)
  \mathcal{T}_{fi}, 
\end{align}
where $P_i$ and $P_f$ stand for the total four momenta of the initial
and final states, respectively. The transition amplitudes
$\mathcal{T}_{fi}$ can be derived from the Bethe-Salpeter integral
equation  
\begin{align}
\mathcal{T}_{fi} (p',p;s) =\, \mathcal{V}_{fi}(p',p;s) 
+ \frac{1}{(2\pi)^4}\sum_k \int d^4q 
\mathcal{V}_{fk}(p',q;s)\mathcal{G}_{k}(q;s) \mathcal{T}_{ki}(q,p;s),  
\label{eq:2}
\end{align}
where $p$ and $p'$ denote the relative four-momenta of the
initial and final states, respectively. $q$ is the off-mass-shell
momentum for the intermediate states in the center of mass (CM) 
frame. $s$ represents the square of the total energy, which is just
one of the Mandelstam variables, $s=P_i^2=P_f^2$. The coupled integral
equations given in Eq.~\eqref{eq:2} can be depicted as in 
Fig.~\ref{fig:1}.
To avoid the complexity due to the four-dimensional integral equations,
we make a three-dimensional reduction. While there are several
different methods for the three-dimensional reduction, we employ the 
Blankenbecler-Sugar scheme~\cite{Blankenbecler:1965gx, Aaron:1968aoz},
which takes the two-body propagator in the form of 
\begin{align}
  \mathcal{G}_k(q) =\;
  \delta\left(q_0-\frac{E_{k1}(\bm{q})-E_{k2}(\bm{q})}{2}\right)
  \frac{\pi}{E_{k1}(\bm{q})E_{k2}(\bm{q})}
  \frac{E_k(\bm{q})}{s-E_k^2(\bm{q})},  
\label{eq:4}
\end{align}
where $E_k$ represents the total on-mass-shell energy of the
intermediate state, $E_k = E_{k1}+E_{k2}$, and $\bm{q}$
denotes the three-momentum of the intermediate state. 
Note that the spinor factors from the meson-baryon
propagator $G_k$ have been absorbed to the matrix elements of 
$\mathcal{V}$ and $\mathcal{T}$. Utilizing Eq.~\eqref{eq:4}, we obtain
the following coupled integral equations   
\begin{align}
  \mathcal{T}_{fi} (\bm{p}',\bm{p}) =\, \mathcal{V}_{fi}
  (\bm{p}',\bm{p}) 
  +\frac{1}{(2\pi)^3}\sum_k\int \frac{d^3q}{2E_{k1}(\bm{q})E_{k2}
  (\bm{q})} \mathcal{V}_{fk}(\bm{p}',\bm{q})\frac{E_k
  (\bm{q})}{s-E_k^2(\bm{q})+i\varepsilon} 
  \mathcal{T}_{ki}(\bm{q},\bm{p}),
  \label{eq:BS-3d}
\end{align}
where $\bm{p}$ and $\bm{p}'$ are the relative three-momenta
of the initial and final states in the CM frame,
respectively.
\begin{figure}[htbp]
  \centering
  \includegraphics[scale=0.35]{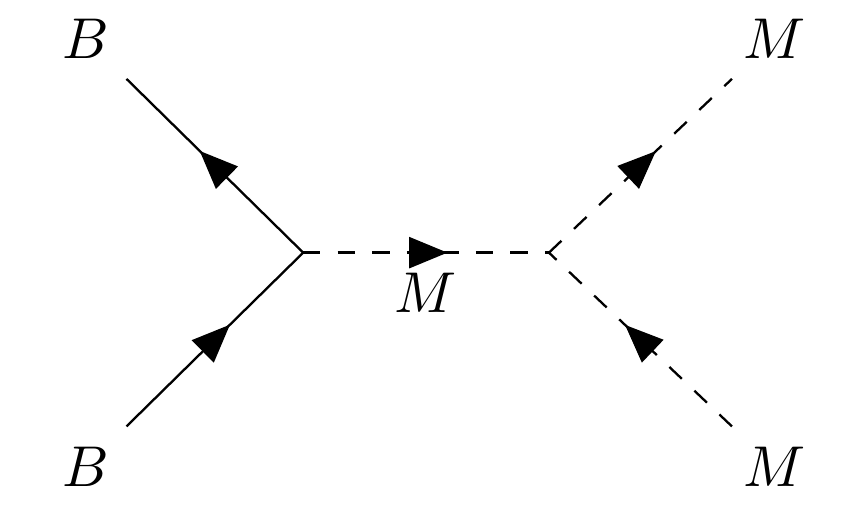}
  \caption{$t$-channel diagrams for the
          meson-exchanged diagrams. $M$ and $B$ stand for the
          meson and baryon, respectively.} 
  \label{fig:2}
\end{figure}
We construct two-body coupled channels by combining the charmed meson
triplet and singly charmed baryon antitriplet and sextet with total
strangeness number $S=0$ to study the pentaquark $P_{c\bar{c}}$.  
In addition, we also introduce the $J/\psi N$ channel, since $P_c$'s
were experimentally known to decay into $J/\psi N$. Thus, we
have the seven different channels as follows: $J/\psi N$, $\bar{D} 
\Lambda_c$, $\bar{D}^*\Lambda_c$, $\bar{D}\Sigma_c$, $\bar{D}
\Sigma_c^*$, $\bar{D}^*\Sigma_c$ and $\bar{D}^*\Sigma_c^*$.
Thus, we first construct the kernel matrix expressed as 
\begin{align}
\mathcal{V} &= \begin{pmatrix}
  \mathcal{V}_{J/\psi N\to J/\psi N} &
        \mathcal{V}_{\bar{D}\Lambda_c\to J/\psi N} &
        \mathcal{V}_{\bar{D}^*\Lambda_c\to J/\psi N} &  
    \mathcal{V}_{\bar{D}\Sigma_c\to J/\psi N} &
    \mathcal{V}_{\bar{D}\Sigma_c^*\to J/\psi N} &
    \mathcal{V}_{\bar{D}^*\Sigma_c\to J/\psi N} & 
    \mathcal{V}_{\bar{D}^*\Sigma_c^*\to J/\psi N}\\ 
  \mathcal{V}_{J/\psi N\to\bar{D}\Lambda_c} &
        \mathcal{V}_{\bar{D}\Lambda_c\to\bar{D}\Lambda_c} &
        \mathcal{V}_{\bar{D}^*\Lambda_c\to\bar{D}\Lambda_c} &  
    \mathcal{V}_{\bar{D}\Sigma_c\to\bar{D}\Lambda_c} &
    \mathcal{V}_{\bar{D}\Sigma_c^*\to\bar{D}\Lambda_c} &
    \mathcal{V}_{\bar{D}^*\Sigma_c\to\bar{D}\Lambda_c} & 
    \mathcal{V}_{\bar{D}^*\Sigma_c^*\to\bar{D}\Lambda_c}\\
  \mathcal{V}_{J/\psi N\to \bar{D}^*\Lambda_c} &
        \mathcal{V}_{\bar{D}\Lambda_c\to \bar{D}^*\Lambda_c} &
        \mathcal{V}_{\bar{D}^*\Lambda_c\to \bar{D}^*\Lambda_c} & 
    \mathcal{V}_{\bar{D}\Sigma_c\to \bar{D}^*\Lambda_c} &
    \mathcal{V}_{\bar{D}\Sigma_c^*\to \bar{D}^*\Lambda_c} &
    \mathcal{V}_{\bar{D}^*\Sigma_c\to \bar{D}^*\Lambda_c} & 
    \mathcal{V}_{\bar{D}^*\Sigma_c^*\to \bar{D}^*\Lambda_c}\\ 
  \mathcal{V}_{J/\psi N\to \bar{D}\Sigma_c} &
        \mathcal{V}_{\bar{D}\Lambda_c\to \bar{D}\Sigma_c}
        &\mathcal{V}_{\bar{D}^*\Lambda_c\to \bar{D}\Sigma_c} &  
    \mathcal{V}_{\bar{D}\Sigma_c\to \bar{D}\Sigma_c} &
    \mathcal{V}_{\bar{D}\Sigma_c^*\to \bar{D}\Sigma_c} &
    \mathcal{V}_{\bar{D}^*\Sigma_c\to \bar{D}\Sigma_c} & 
    \mathcal{V}_{\bar{D}^*\Sigma_c^*\to \bar{D}\Sigma_c}\\
  \mathcal{V}_{J/\psi N\to \bar{D}\Sigma_c^*} &
        \mathcal{V}_{\bar{D}\Lambda_c\to \bar{D}\Sigma_c^*} &
        \mathcal{V}_{\bar{D}^*\Lambda_c\to \bar{D}\Sigma_c^*} &  
    \mathcal{V}_{\bar{D}\Sigma_c\to \bar{D}\Sigma_c^*} &
    \mathcal{V}_{\bar{D}\Sigma_c^*\to \bar{D}\Sigma_c^*} &
    \mathcal{V}_{\bar{D}^*\Sigma_c\to \bar{D}\Sigma_c^*} & 
    \mathcal{V}_{\bar{D}^*\Sigma_c^*\to \bar{D}\Sigma_c^*}\\
  \mathcal{V}_{J/\psi N\to \bar{D}^*\Sigma_c} &
        \mathcal{V}_{\bar{D}\Lambda_c\to \bar{D}^*\Sigma_c} &
        \mathcal{V}_{\bar{D}^*\Lambda_c\to \bar{D}^*\Sigma_c} & 
    \mathcal{V}_{\bar{D}\Sigma_c\to \bar{D}^*\Sigma_c} &
    \mathcal{V}_{\bar{D}\Sigma_c^*\to \bar{D}^*\Sigma_c} &
    \mathcal{V}_{\bar{D}^*\Sigma_c\to \bar{D}^*\Sigma_c} & 
    \mathcal{V}_{\bar{D}^*\Sigma_c^*\to \bar{D}^*\Sigma_c}\\
  \mathcal{V}_{J/\psi N\to \bar{D}^*\Sigma_c^*} &
        \mathcal{V}_{\bar{D}\Lambda_c\to \bar{D}^*\Sigma_c^*} &
        \mathcal{V}_{\bar{D}^*\Lambda_c\to \bar{D}^*\Sigma_c^*} & 
    \mathcal{V}_{\bar{D}\Sigma_c\to \bar{D}^*\Sigma_c^*} &
    \mathcal{V}_{\bar{D}\Sigma_c^*\to \bar{D}^*\Sigma_c^*} &
    \mathcal{V}_{\bar{D}^*\Sigma_c\to \bar{D}^*\Sigma_c^*} &  
    \mathcal{V}_{\bar{D}^*\Sigma_c^*\to \bar{D}^*\Sigma_c^*}\\
  \end{pmatrix}.
    \label{eq:kernel}
\end{align}
Each component of the kernel matrix is constructed by using one-meson 
exchange tree-level diagram which is shown in Fig.~\ref{fig:2}.
We do not include pole diagrams in the $s$ channel.  
We consider the $t$-channel diagrams, which are essential for
generating the $P_{c\bar{c}}$ states dynamically. The contributions of
$u$-channel diagrams are very small, so we ignore them. The
interactions at the vertex are governed by the effective Lagrangian
that respects heavy-quark spin symmetry, hidden local gauge symmetry,
and flavor SU(3) symmetry~\cite{Casalbuoni:1996pg}. The mesonic
vertices are then computed by the effective Lagrangian given as  
\begin{align}
  \mathcal{L}_{PP\mathbb{V}} &= -i\frac{\beta g_V}{\sqrt{2}}\,
  P^{\dagger}_a  \overleftrightarrow{\partial_\mu} P_b\,
   \mathbb{V}^\mu_{ba} +i\frac{\beta g_V}{\sqrt{2}}\,P'^{\dagger}_a
                               \overleftrightarrow{\partial_\mu}
                               P'_b\, \mathbb{V}^\mu_{ab},\\ 
    \mathcal{L}_{PP\sigma} &= -2g_\sigma M P^\dagger_a \sigma P_a
  -2g_\sigma M P'^\dagger_a\sigma P'_a ,\\     %
    \mathcal{L}_{P^*P^*\mathbb{P}} &= -\frac{g}{f_\pi}
  \epsilon^{\mu\nu\alpha\beta}P^{*\dagger}_{a\nu}\,
   \overleftrightarrow{\partial_\mu}\,
P^*_{b\beta}\partial_\alpha \mathbb{P}_{ba} -\frac{g}{f_\pi}
  \epsilon^{\mu\nu\alpha\beta}  P'^{*\dagger}_{a\nu}\,
 \overleftrightarrow{\partial_\mu}\,
 P'^*_{b\beta}\partial_\alpha
                                     \mathbb{P}_{ab} ,\\     %
    \mathcal{L}_{P^*P^*\mathbb{V}} & = i\frac{\beta g_V}{\sqrt{2}} \,
 P^{*\dagger}_{a\nu}  \overleftrightarrow{\partial_\mu}
 P^{*\nu}_b \mathbb{V}_{ba}^\mu + i2\sqrt{2} \lambda g_VM^*
 P^{*\dagger}_{a\mu}  P^*_{b\nu}\mathbb{V}_{ba}^{\mu\nu}\cr
  &\;\;\;\;-i\frac{\beta g_V}{\sqrt{2}} \,P'^{*\dagger}_{a\nu}
    \overleftrightarrow{\partial_\mu} P'^{*\nu}_b
    \mathbb{V}_{ab}^\mu-i2\sqrt{2}\lambda g_VM^*
    P'^{*\dagger}_{a\mu} P'^*_{b\nu}\mathbb{V}_{ab}^{\mu\nu} ,\\
  \mathcal{L}_{P^*P^*\sigma} &= 2g_\sigma M^* P^{*\dagger}_{a\mu}
\sigma P^{*\mu}_a+2g_\sigma M^*
 P'^{*\dagger}_{a\mu}\sigma P'^{*\mu}_a ,\\    %
    \mathcal{L}_{P^*P\mathbb{P}} &= -\frac{2g}{f_\pi}  \sqrt{MM^*}\,
      \left( P^{\dagger}_a P^*_{b\mu}+P^{*\dagger}_{a\mu} P_b\right)\,
 \partial^\mu \mathbb{P}_{ba}+\frac{2g}{f_\pi} \sqrt{MM^*}\,
    \left(P'^{\dagger}_a P'^*_{b\mu}+P'^{*\dagger}_{a\mu} P'_b\right)\,
  \partial^\mu \mathbb{P}_{ab},\\
  \mathcal{L}_{P^*P\mathbb{V}} &= -i\sqrt{2}\lambda g_V\,
\epsilon^{\beta\alpha\mu\nu} \left(P^{\dagger}_a
\overleftrightarrow{\partial_\beta} P^*_{b\alpha} +
P^{*\dagger}_{a\alpha}  \overleftrightarrow{\partial_\beta}
  P_b\right)\,\left(\partial_\mu\mathbb{V}_{\nu}\right)_{ba}\cr 
 &\;\;\;\;-i\sqrt{2}\lambda g_V\, \epsilon^{\beta\alpha\mu\nu}
   \left(P'^{\dagger}_a\overleftrightarrow{\partial_\beta}
   P'^*_{b\alpha}+P'^{*\dagger}_{a\alpha}
   \overleftrightarrow{\partial_\beta}P'_b\right)\,
   \left(\partial_\mu\mathbb{V}_{\nu}\right)_{ab}.
\end{align}
with $\overleftrightarrow{\partial} =
\overrightarrow{\partial}-\overleftarrow{\partial}$. The lowest
isoscalar-scalar meson is denoted by $\sigma$. The heavy meson and
anti heavy meson matrices $P^{(*)}$ and $P'^{(*)}$ are given by 
\begin{align}
  P = \left(D^0,D^+,D_s^+\right), \hspace{0.5 cm}
  P^*_\mu =\left(D^{*0}_\mu,D^{*+}_\mu,D_{s\mu}^{*+}\right),
  \hspace{0.5 cm} P' =(\bar{D}^0,\,D^-,\,D_s^-),
  \hspace{0.5 cm} P'^*_\mu =(\bar{D}^{*0}_\mu,\,D^{*-}_\mu,\,D^{*-}_{s\mu}),
\end{align}
while the light pseudoscalar and vector meson matrices are
\begin{align}
    \mathbb{P} = 
    \begin{pmatrix}
        \frac{1}{\sqrt{2}} \pi^0+\frac{1}{\sqrt{6}}\eta & \pi^+ & K^+\\
        \pi^- & -\frac{1}{\sqrt{2}} \pi^0+\frac{1}{\sqrt{6}}\eta & K^0\\
        K^- & \bar{K}^0 & -\frac{2}{\sqrt{6}}\eta
    \end{pmatrix},\;\;\;\;
    \mathbb{V}_\mu = \begin{pmatrix}
        \frac{1}{\sqrt{2}} \rho^0_\mu+\frac{1}{\sqrt{2}}\omega_\mu &
        \rho_\mu^+ & K_\mu^{*+}\\
        \rho_\mu^- & -\frac{1}{\sqrt{2}} \rho_\mu^0+\frac{1}{\sqrt{2}}
        \omega_\mu & K_\mu^{*0} \\
        K_\mu^{*-} & \bar{K}^{*0}_\mu & \phi_\mu
    \end{pmatrix}.
\end{align}
The coupling constants in the Lagrangian are obtained from
Ref.~\cite{Isola:2003fh}, i.e., $g=0.59\pm0.07\pm0.01$ from
experimental results of $D^{*+}$ full width, $g_V=m_\rho/f_\pi\approx
5.8$ by using the KSRF relation with $f_\pi=132$ MeV, $\beta\approx
0.9$ by assuming vector meson dominance in the radiative decay of
heavy mesons, and $\lambda=-0.56\,\mathrm{GeV}^{-1}$ by using
light-cone sum rules and lattice QCD. Notice that we use a different
sign of $\lambda$ from Ref.~\cite{Isola:2003fh}, since we use the same
phase of heavy vector meson as in Ref.~\cite{Casalbuoni:1996pg}. The
coupling constant for the sigma meson is utilized to calculate the $2\pi$ 
transition of $D_s(1^+)$ in Ref.~\cite{Bardeen:2003kt}. The lowest
isoscalar-scalar meson coupling is $g_\sigma=g_\pi/2\sqrt{6}$ with
$g_\pi=3.73$. 

As for the effective Lagrangian for the heavy baryon, we take it from 
Ref.~\cite{Liu:2011xc}, where a more general form of the Lagrangian
was considered~\cite{Yan:1992gz}. The interaction vertices for the
baryonic sector in the tree level diagram of the meson-exchange
diagram are governed by the following effective Lagrangian 
\begin{align}
    \mathcal{L}_{B_{\bar{3}}B_{\bar{3}}\mathbb{V}}&= \frac{i\beta_{\bar{3}}g_V}{2\sqrt{2}M_{\bar{3}}}\left(\bar{B}_{\bar{3}}\overleftrightarrow{\partial_\mu} \mathbb{V}^\mu B_{\bar{3}}\right) ,\\
    \mathcal{L}_{B_{\bar{3}}B_{\bar{3}}\sigma}&=l_{\bar{3}}\left(\bar{B}_{\bar{3}}\sigma B_{\bar{3}}\right) ,\\
    \mathcal{L}_{B_6B_6\mathbb{P}}&= i\frac{g_1}{2f_\pi M_6}\bar{B}_{6}\gamma_5\left(\gamma^\alpha\gamma^\beta-g^{\alpha\beta}\right)\overleftrightarrow{\partial_\alpha}\partial_\beta\mathbb{P} B_{6} ,\\
    \mathcal{L}_{B_6B_6\mathbb{V}}&= -i\frac{\beta_6 g_V}{2\sqrt{2}M_6}\left(\bar{B}_{6}\overleftrightarrow{\partial_\alpha}\mathbb{V}^\alpha B_{6}\right)-\frac{i\lambda_6g_V}{3\sqrt{2}}\left(\bar{B}_{6}\gamma_\mu\gamma_\nu \mathbb{V}^{\mu\nu}B_{6}\right) ,\\
    \mathcal{L}_{B_6B_6\sigma}&= -l_6\left(\bar{B}_{6}\sigma B_6\right) ,\\
    \mathcal{L}_{B_6^*B_6^*\mathbb{P}}&= \frac{3g_1}{4f_\pi M_6^*}\epsilon^{\mu\nu\alpha\beta}\left(\bar{B}_{6\mu}^*\overleftrightarrow{\partial_\nu}\partial_\alpha\mathbb{P} B_{6\beta}^*\right) ,\\
    \mathcal{L}_{B_6^*B_6^*\mathbb{V}}&= i\frac{\beta_6 g_V}{2\sqrt{2}M_6^*}\left(\bar{B}_{6\mu}^*\overleftrightarrow{\partial_\alpha}\mathbb{V}^\alpha B_{6}^{*\mu}\right)+\frac{i\lambda_6g_V}{\sqrt{2}} \left(\bar{B}^*_{6\mu} \mathbb{V}^{\mu\nu}B^*_{6\nu}\right) ,\\
    \mathcal{L}_{B_6^*B_6^*\sigma}&= l_6\left(\bar{B}^*_{6\mu}\sigma B^{*\mu}_6\right) ,\\
    \mathcal{L}_{B_6B_6^*\mathbb{P}}&= \frac{g_1}{4f_\pi}\sqrt{\frac{3}{M_6^*M_6}}\epsilon^{\mu\nu\alpha\beta}\left[\left(\bar{B}_{6}\gamma_5\gamma_\mu\overleftrightarrow{\partial_\nu} \partial_\alpha\mathbb{P} B_{6\beta}^*\right)+\left(\bar{B}_{6\mu}^*\gamma_5\gamma_\nu\overleftrightarrow{\partial_\alpha} \partial_\beta\mathbb{P}B_6 \right)\right] ,\\
    \mathcal{L}_{B_6B_6^*\mathbb{V}}&= \frac{i\lambda_6g_V}{\sqrt{6}}\left[\bar{B}_{6}\gamma_5\left(\gamma_\mu +\frac{i\overleftrightarrow{\partial_\mu}}{2\sqrt{M_6^*M_6}}\right) \mathbb{V}^{\mu\nu}B^*_{6\nu}+\bar{B}_{6\mu}^*\gamma_5\left(\gamma_\nu -\frac{i\overleftrightarrow{\partial_\nu}}{2\sqrt{M_6^*M_6}}\right) \mathbb{V}^{\mu\nu}B_{6}\right] ,\\
    \mathcal{L}_{B_6B_{\bar{3}}\mathbb{P}}&= -\frac{g_4}{\sqrt{3}f_\pi}\left[\bar{B}_6\gamma_5\left(\gamma_\mu +\frac{i\overleftrightarrow{\partial_\mu}}{2\sqrt{M_6M_{\bar{3}}}} \right) \partial^\mu\mathbb{P} B_{\bar{3}}+\bar{B}_{\bar{3}}\gamma_5\left(\gamma_\mu -\frac{i\overleftrightarrow{\partial_\mu}}{2\sqrt{M_6M_{\bar{3}}}} \right) \partial^\mu\mathbb{P}\,B_{6}\right] ,\\
    \mathcal{L}_{B_6B_{\bar{3}}\mathbb{V}}&= i\frac{\lambda_{6\bar{3}}\,g_V}{\sqrt{6M_6M_{\bar{3}}}}\epsilon^{\mu\nu\alpha\beta}\left[\left(\bar{B}_{6}\gamma_5\gamma_\mu \overleftrightarrow{\partial_\nu}\partial_\alpha\mathbb{V}_{\beta} B_{\bar{3}}\right)+\left(\bar{B}_{\bar{3}}\gamma_5\gamma_\mu \overleftrightarrow{\partial_\nu}\partial_\alpha\mathbb{V}_{\beta} B_{6}\right)\right] ,\\
    \mathcal{L}_{B_6^*B_{\bar{3}}\mathbb{P}}&= -\frac{g_4}{f_\pi}\left[\left(\bar{B}^*_{6\mu}\partial^\mu\mathbb{P} B_{\bar{3}}\right)+\left(\bar{B}_{\bar{3}}\partial^\mu\mathbb{P} B^*_{6\mu}\right)\right] ,\\
    \mathcal{L}_{B_6^*B_{\bar{3}}\mathbb{V}}&= i\frac{\lambda_{6\bar{3}}\,g_V}{\sqrt{2M_6^*M_{\bar{3}}}}\epsilon^{\mu\nu\alpha\beta}\left[\left(\bar{B}^*_{6\mu}\overleftrightarrow{\partial_\nu}\partial_\alpha\mathbb{V}_{\beta} B_{\bar{3}}\right)+\left(\bar{B}_{\bar{3}}\overleftrightarrow{\partial_\nu}\partial_\alpha\mathbb{V}_{\beta} B^*_{6\mu}\right)\right] ,
\end{align}
where the heavy baryon fields are given by
\begin{align}
    &B_{\bar{3}}=
    \begin{pmatrix}
        0 & \Lambda_c^+ & \Xi_c^+\\
        -\Lambda_c^+ & 0 & \Xi_c^0\\
        -\Xi_c^+ & -\Xi_c^0 & 0
    \end{pmatrix},\;\;
    B_6 =
    \begin{pmatrix}
        \Sigma_c^{++} & \frac{1}{\sqrt{2}}\Sigma_c^+ & \frac{1}{\sqrt{2}}\Xi{'}_c^+\\
        \frac{1}{\sqrt{2}}\Sigma_c^+ & \Sigma_c^0 & \frac{1}{\sqrt{2}}\Xi{'}_c^0\\
        \frac{1}{\sqrt{2}}\Xi{'}_c^+ & \frac{1}{\sqrt{2}}\Xi{'}_c^0 & \Omega_c^0
    \end{pmatrix},\;\;
    B_6^* =
    \begin{pmatrix}
        \Sigma_c^{*++} & \frac{1}{\sqrt{2}}\Sigma_c^{*+} & \frac{1}{\sqrt{2}}\Xi_c^{*+}\\
        \frac{1}{\sqrt{2}}\Sigma_c^{*+} & \Sigma_c^{*0} & \frac{1}{\sqrt{2}}\Xi_c^{*0}\\
        \frac{1}{\sqrt{2}}\Xi_c^{*+} & \frac{1}{\sqrt{2}}\Xi_c^{*0} & \Omega_c^{*0}
    \end{pmatrix}.
\end{align}
$B_\mu$ denotes the spin 3/2 Rarita-Schwinger field, which satisfies
the following constraint
\begin{align}
  p^\mu B_\mu = 0 \hspace{0.5 cm}{\rm and}\hspace{0.5 cm}
  \gamma^\mu B_\mu = 0 .
\end{align}
The coupling constants in the effective Lagrangian are given as
follows~\cite{Liu:2011xc,Chen:2019asm}: $\beta_{\bar{3}} = 6/g_V$,
$\beta_6=-2\beta_{\bar{3}}$, $\lambda_6=-3.31\,\mathrm{GeV}^{-1}$,
$\lambda_{6\bar{3}}=-\lambda_6/\sqrt{8}$, $g_1=0.942$ and $g_4=0.999$.
The different signs used above are taken from
Refs.~\cite{Chen:2019asm, Dong:2021juy}. 

Since we include the hidden-charm channel, we need effective
Lagrangian to describe the coupling between heavy mesons and
quarkonium. Here we use the Lagrangian from
Ref.~\cite{Colangelo:2003sa}, i.e., 
\begin{align}
    \mathcal{L}_{PPJ/\psi} &= -ig_\psi M\sqrt{m_{J}}\left(J/\psi^\mu P^\dagger\overleftrightarrow{\partial_\mu}P{'}^{\dagger}\right) + \mathrm{h.c.,}\\
    \mathcal{L}_{P^*PJ/\psi} &= ig_\psi\sqrt{\frac{MM^*}{m_{J}}}\epsilon^{\mu\nu\alpha\beta} \partial_\mu J/\psi_\nu\left(P^\dagger\overleftrightarrow{\partial_\alpha}P^*{'}^\dagger_\beta+P_{\beta}^{*\dagger}\overleftrightarrow{\partial_\alpha}P{'}^{\dagger}\right)+\mathrm{h.c.,}\\
    \mathcal{L}_{P^*P^*J/\psi} &= ig_\psi M^*\sqrt{m_J}(g^{\mu\nu}g^{\alpha\beta}-g^{\mu\alpha}g^{\nu\beta}+g^{\mu\beta}g^{\nu\alpha}) \left(J/\psi_\mu P_{\nu}^{*\dagger}\overleftrightarrow{\partial_\alpha}P^*{'}^\dagger_\beta\right)+\mathrm{h.c.}
\end{align}
In this work, we only consider the vector quarkonia since it is
directly related to experiments. However, the extension to the
pseudoscalar state is straightforward since we assume the heavy quark
spin symmetry to the quarkonia state as
well~\cite{Casalbuoni:1992fd}. Since there is no experimental data on
the $J/\psi\to D\bar{D}$ decay, Shimizu et al.~\cite{Shimizu:2017xrg}
estimated the value of the coupling constant $g_\psi$ as follows:
the coupling constant $g_{\phi K\bar{K}}$ can be determined from
the experimental decay width for the $\phi\to K\bar{K}$ decay. 
Assuming that the decay of $J/\psi$ is similar to that of $\phi$ apart
from their masses, Shimizu et al. estimated $g_\psi$ to be
$g_\psi=0.679\,\mathrm{GeV}^{-3/2}$.  The coupling constants of the
heavy baryons and heavy mesons are expressed~\cite{Shimizu:2017xrg} as    
\begin{align}
    \mathcal{L}_{B_8B_3P} &=
   -g_{I{\bar{3}}}\sqrt{M}\bar{B}_{\bar{3}}\gamma_5
   P N+\mathrm{h.c.},\\ 
    \mathcal{L}_{B_8B_3P^*} &=
    g_{I{\bar{3}}}\sqrt{M^*}\bar{B}_{\bar{3}}\gamma^\mu
    P_\mu^* N+\mathrm{h.c.,}\\ 
    \mathcal{L}_{B_8B_6P}&= g_{I6} \sqrt{3M} \bar{B}_{6}\gamma_5 B_8 P
        + \mathrm{h.c.,}\\ 
    \mathcal{L}_{B_8B_6P^*}&= g_{I6}\sqrt{\frac{M^*}{3}}
   \bar{B}_{6}\gamma^\nu B_8 P^*_\nu +
   \mathrm{h.c.,}\\ 
    \mathcal{L}_{B_8B_6^*P^*}&= 2 g_{I6}\sqrt{M^*}
    \bar{B}_{6}^\mu\gamma_5 B_8 P^*_\mu +
    \mathrm{h.c.}. 
\end{align}
We employ the coupling constants 
$g_{I\bar{3}}=-9.88\,\mathrm{GeV}^{-1/2}$ and 
$g_{I6}=1.14\,\mathrm{GeV}^{-1/2}$ taken from
Ref.~\cite{Shimizu:2017xrg}. It is important to note that the coupling
to the hidden charm channels have only a marginal effect to the
production mechanism of the resonance. The current calculation implies
that though these values of the coupling constants are taken from the
rough estimation, the predicted masses of the hidden charm pentaquarks 
almost do not vary. This already indicates that the $J/\psi N$ channel
has a tiny effect on the production of the heavy pentaquarks.

The Feynman amplitude for one-meson exchange diagram can
be written as  
\begin{align}
  \mathcal{A}_{\lambda'_1\lambda'_2,\lambda_1\lambda_2}
  = \mathrm{IS} \,F^2(q^2)\,\Gamma_{\lambda'_1\lambda'_2}(p'_1,p'_2)
  \mathcal P(q)\Gamma_{\lambda_1\lambda_2}(p_1,p_2) ,
\end{align}
where $\lambda_i$ and $p_i$ denote the helicity and momentum of the
corresponding particle respectively, while $q$ is the momentum of the
exchange particle. The IS factor is related to the SU(3)
Clebsch-Gordan coefficient and isospin factor. The IS factor for each
exchanged diagram is listed in Table~\ref{tab:1}.
\begin{table}[htbp]
  \caption{\label{tab:1}The values of the IS factors and
          $\Lambda-m$ for the corresponding $t$-channel diagrams for
          the given reactions. The $\Lambda$ denotes the cutoff mass
          and $m$ stands for the mass of the exchanged particle, given
          in units of MeV.  
        } 
  \begin{ruledtabular}
  \centering\begin{tabular}{lccr}
   \multirow{2}{*}{Reactions} & \multirow{2}{*}{Exchange particles} & 
   \multirow{2}{*}{IS} &   \multirow{2}{*}{$\Lambda-m$} \\
   & & & 
   \\\hline
     $J/\psi N\to\bar{D}\Lambda_c$ 
     & $\bar{D}$, $\bar{D}^*$  & $1$ & $500$ \\
     $J/\psi N\to\bar{D}^*\Lambda_c$ 
     & $\bar{D}$, $\bar{D}^*$  & $1$ & $500$ \\
     $J/\psi N\to\bar{D}\Sigma_c$ 
     & $\bar{D}$, $\bar{D}^*$ & $-\sqrt{3/2}$ & $500$ \\
     $J/\psi N\to\bar{D}\Sigma_c^*$ 
     & $\bar{D}$, $\bar{D}^*$ & $-\sqrt{3/2}$ & $500$ \\
     $J/\psi N\to\bar{D}^*\Sigma_c$ 
     & $\bar{D}$, $\bar{D}^*$ & $-\sqrt{3/2}$ & $500$ \\
     $J/\psi N\to\bar{D}^*\Sigma_c^*$ 
     & $\bar{D}$, $\bar{D}^*$ & $-\sqrt{3/2}$ & $500$ \\
     $\bar{D}\Lambda_c\to\bar{D}\Lambda_c$ 
    & $\omega$ & $1$ & $500$ \\
    & $\sigma$ & $2$ & $500$ \\
     $\bar{D}\Lambda_c\to\bar{D}^*\Lambda_c$ 
    & $\omega$ & $1$ & $500$ \\
    $\bar{D}\Lambda_c\to\bar{D}\Sigma_c$ 
    & $\rho$ & $-\sqrt{3/2}$ & $500$ \\
    $\bar{D}\Lambda_c\to\bar{D}\Sigma_c^*$ 
    & $\pi$, $\rho$ & $-\sqrt{3/2}$ & $500$ \\
    $\bar{D}\Lambda_c\to\bar{D}^*\Sigma_c$ 
    & $\pi$, $\rho$ & $-\sqrt{3/2}$ & $500$ \\
    $\bar{D}\Lambda_c\to\bar{D}^*\Sigma_c^*$ 
    & $\pi$, $\rho$ & $-\sqrt{3/2}$ & $500$ \\
     $\bar{D}^*\Lambda_c\to\bar{D}^*\Lambda_c$ 
    & $\omega$ & $1$ & $500$ \\
    & $\sigma$ & $2$ & $500$ \\
    $\bar{D}^*\Lambda_c\to\bar{D}\Sigma_c$ 
    & $\pi$, $\rho$ & $-\sqrt{3/2}$ & $500$ \\
    $\bar{D}^*\Lambda_c\to\bar{D}\Sigma_c^*$ 
    & $\pi$, $\rho$ & $-\sqrt{3/2}$ & $500$ \\
    $\bar{D}^*\Lambda_c\to\bar{D}^*\Sigma_c$ 
    & $\pi$, $\rho$ & $-\sqrt{3/2}$ & $500$ \\
    $\bar{D}^*\Lambda_c\to\bar{D}^*\Sigma_c^*$ 
    & $\pi$, $\rho$ & $-\sqrt{3/2}$ & $500$ \\
    $\bar{D}\Sigma_c\to\bar{D}\Sigma_c$ 
    & $\rho$ & $-1$ & $500$ \\
    & $\omega$ & $1/2$ & $500$ \\
    & $\sigma$ & $1$ & $500$ \\
    $\bar{D}\Sigma_c\to\bar{D}\Sigma_c^*$ 
    & $\rho$ & $-1$ & $500$ \\
    & $\omega$ & $1/2$ & $500$ \\
    $\bar{D}\Sigma_c\to\bar{D}^*\Sigma_c$ 
    & $\pi$, $\rho$ & $-1$ & $500$ \\
    & $\eta$ & $1/6$ & $500$ \\
    & $\omega$ & $1/2$ & $500$ \\
    $\bar{D}\Sigma_c\to\bar{D}^*\Sigma_c^*$ 
    & $\pi$, $\rho$ & $-1$ & $500$ \\
    & $\eta$ & $1/6$ & $500$ \\
    & $\omega$ & $1/2$ & $500$ \\
    $\bar{D}\Sigma_c^*\to\bar{D}\Sigma_c^*$ 
    & $\rho$ & $-1$ & $700$ \\
    & $\omega$ & $1/2$ & $700$ \\
    & $\sigma$ & $1$ & $700$ \\
    $\bar{D}\Sigma_c^*\to\bar{D}^*\Sigma_c$ 
    & $\pi$, $\rho$ & $-1$ & $700$ \\
    & $\eta$ & $1/6$ & $700$ \\
    & $\omega$ & $1/2$ & $700$ \\
    $\bar{D}\Sigma_c^*\to\bar{D}^*\Sigma_c^*$ 
    & $\pi$, $\rho$ & $-1$ & $700$ \\
    & $\eta$ & $1/6$ & $700$ \\
    & $\omega$ & $1/2$ & $700$ \\
    $\bar{D}^*\Sigma_c\to\bar{D}^*\Sigma_c$ 
    & $\pi$, $\rho$ & $-1$ & $700$ \\
    & $\eta$ & $1/6$ & $700$ \\
    & $\omega$ & $1/2$ & $700$ \\
    & $\sigma$ & $1$ & $700$ \\
    $\bar{D}^*\Sigma_c\to\bar{D}^*\Sigma_c^*$ 
    & $\pi$, $\rho$ & $-1$ & $700$ \\
    & $\eta$ & $1/6$ & $700$ \\
    & $\omega$ & $1/2$ & $700$ \\
    $\bar{D}^*\Sigma_c^*\to\bar{D}^*\Sigma_c^*$ 
    & $\pi$, $\rho$ & $-1$ & $700$ \\
    & $\eta$ & $1/6$ & $700$ \\
    & $\omega$ & $1/2$ & $700$ \\
    & $\sigma$ & $1$ & $700$ \\
  \end{tabular}
    \end{ruledtabular}
\end{table}
The vertex $\Gamma$ is derived by using the effective Lagrangian
previously described and the propagators for the spin-0 and spin-1
mesons are expressed as     
\begin{align}
  \mathcal{P}(q) &= \frac{1}{q^2-m^2},\;\;\;
  \mathcal{P}_{\mu\nu}(q) = \frac{1}{q^2-m^2}
  \left(-g_{\mu\nu}+\frac{q_\mu q_\nu}{m^2}\right).
\end{align}
We use the static propagator for pion exchange,
$\mathcal{P}_\pi(q)=-1/(\bm{q}^2 +m_\pi^2)$ for simplicity. As for the
heavy-meson propagators, we employ the same form as the light mesons,
since the heavy-quark mass is actually finite.  
The parity invariance further reduces a number of proceseses. The
parity relation is given by
\begin{align}
  \mathcal{A}_{-\lambda'_1-\lambda'_2,-\lambda_1-\lambda_2} =
  \eta(\eta')^{-1}\,
  \mathcal{A}_{\lambda'_1\lambda'_2,\lambda_1\lambda_2},  
    \label{eq:ampi}
\end{align}
where $\eta$ is expressed as 
\begin{align}
    \eta = \eta_1\eta_2(-1)^{J-s_1-s_2}.
\end{align}
$\eta_i$ and $s_i$ denote the intrinsic parity and spin of the
particle, respectively, while $J$ designates the total angular
momentum.  

Since hadrons have finite sizes, we introduce a form factor
at each vertex. To this end, we employ the following
parametrization~\cite{Kim:1994ce} 
\begin{align}
  F(q^2) = \left(\frac{n\Lambda^2-m^2}
  {n\Lambda^2-q^2}
  \right)^n,
\label{eq:13}
\end{align}
where $n$ is determined by the power of the momentum in the
vertex. This parametrization has the advantage that we do not need to
adjust the value of $\Lambda$ when we change $n$. It is worth noting
that when we take the limit $n\to \infty$, Eq.~\eqref{eq:13} becomes a
Gaussian form. While the cut-off masses $\Lambda$ in Eq.~\eqref{eq:13}
are not experimentally known for heavy hadron processes, we adopt a 
strategy to minimize the associated uncertainties. We
determine $\Lambda$ by adding approximately $(500-700)$ MeV to the 
corresponding masses of the exchange meson. 
Recent studies have explicitly shown that heavy hadrons are more
compact than light ones~\cite{Kim:2018nqf,
  Kim:2021xpp}. This indicates that the cutoff masses for heavy
hadrons must be larger than those of light ones. 
Consequently, we set the value of cutoff mass as
$\Lambda=\Lambda_0+m$, where $m$ is the mass of the exchange
meson. Thus, we choose $\Lambda_0$ to be approximately $500-700$ MeV
for each channel, as listed in Table.~\ref{tab:1}. This approach 
allows us to perform a minimal fitting procedure.  

To further simplify the numerical calculation and the
spin-parity assignments for the $P_{c\bar{c}}$ states, we carry out a 
partial-wave expansion of the $\mathcal{V}$ and $\mathcal{T}$
matrices. This yields a one-dimensional integral equation given by  
\begin{align}
  \mathcal{T}^{J(fi)}_{\lambda'\lambda} (\mathrm{p}',\mathrm{p}) = 
  \mathcal{V}^{J(fi)}_{
  \lambda'\lambda} (\mathrm{p}',\mathrm{p}) + \frac{1}{(2\pi)^3}
  \sum_{k,\lambda_k}\int
  \frac{\mathrm{q}^2d\mathrm{q}}{2E_{k1}E_{k2}}
  \mathcal{V}^{J(fk)}_{\lambda'\lambda_k}(\mathrm{p}',
  \mathrm{q})\frac{E_k}{
  s-E_k^2+i\varepsilon} \mathcal{T}^{J(ki)}_{\lambda_k\lambda}
  (\mathrm{q},\mathrm{p}),
  \label{eq:BS-1d}
\end{align}
where $\lambda'=\{\lambda'_1,\lambda'_2\}$,
$\lambda=\{\lambda_1,\lambda_2\}$ and
$\lambda_k=\{\lambda_{k1},\lambda_{k2}\}$ denote the helicities of the
final ($f$), initial ($i$) and intermediate ($k$) states,
respectively. The partial-wave kernel amplitudes
$\mathcal{V}_{\lambda'\lambda}^{J(fi)}$ can be 
expressed as 
\begin{equation}
  \mathcal{V}^{J(fi)}_{\lambda'\lambda}(\mathrm{p}',\mathrm{p}) = 
  2\pi \int d( \cos\theta) \,
        d^{J}_{\lambda'_1-\lambda'_2,\lambda_1-\lambda_2}(\theta)\,
        \mathcal{V}^{fi}_{\lambda'\lambda}(\mathrm{p}',\mathrm{p},\theta),
\end{equation} 
where $\theta$ represents the scattering angle and
$d^{J}_{\lambda_f\lambda_i}(\theta)$ denotes the matrix elements of
the Wigner $D$ functions. 

The integral equation in Eq.~\eqref{eq:BS-1d} contains the  
singularity originating from the two-body propagator $\mathcal{G}$. 
To manage this singularity, we isolate its singular part and treat 
it separately. The resulting regularized integral equation is
expressed as 
\begin{align}
  \mathcal{T}^{fi}_{\lambda'\lambda} (\mathrm{p}',\mathrm{p}) = 
  \mathcal{V}^{fi}_{
  \lambda'\lambda} (\mathrm{p}',\mathrm{p}) + \frac{1}{(2\pi)^3}
  \sum_{k,\lambda_k}\left[\int_0^{\infty}d\mathrm{q}
  \frac{\mathrm{q}E_k}{E_{k1}E_{k2}}\frac{\mathcal{F}(\mathrm{q})
  -\mathcal{F}(\tilde{\mathrm{q}}_k)}{s-E_k^2}+ \frac{1}{2\sqrt{s}}
  \left(\ln\left|\frac{\sqrt{s}-E_k^{\mathrm{thr}}}{\sqrt{s}
  +E_k^{\mathrm{thr}}}\right|-i\pi\right)\mathcal{F}
  (\tilde{\mathrm{q}}_k)\right],
  \label{eq:BS-1d-reg}
\end{align}
with 
\begin{align}
  \mathcal{F}(\mathrm{q})=\frac{1}{2}\mathrm{q}\,
  \mathcal{V}^{fk}_{\lambda'\lambda_k}(\mathrm{p}',
  \mathrm{q})\mathcal{T}^{ki}_{\lambda_k\lambda}(\mathrm{q},\mathrm{p}) ,
\end{align}
and $\tilde{\mathrm{q}}_k$ is the momentum $\mathrm{q}$ when
$E_{k1}+E_{k2}=\sqrt{s}$.  
The regularization is applied only when the total energy $\sqrt{s}$ exceeds 
the threshold energy of the $k$-th channel $E_k^{\mathrm{thr}}$. It is
important to note that the form factors introduced in the amplitude
$\mathcal{V}$ provide sufficient suppression in the high-momentum
region, which allows for the regularization of the integration.  

To compute $\mathcal{T}$ from Eq.~\eqref{eq:BS-1d-reg} numerically, we  
expand the $\mathcal{V}$ matrix in helicity states and momentum
space, with momenta obtained by using the Gaussian quadrature
method. We then derive the $\mathcal{T}$ matrix using the
Haftel-Tabakin method for matrix inversion~\cite{Haftel:1970zz}
\begin{align}
  \mathcal{T} = \left(1-\mathcal{V}\tilde{\mathcal{G}}\right)^{-1} 
  \mathcal{V}.
\end{align}
The resulting $\mathcal{T}$ matrix is in the helicity basis and
lacks definite parity. To study parity assignments for 
$P_{c\bar{c}}$, we decompose the transition amplitudes
into the partial-wave amplitudes with definite parity given by 
\begin{align}
  \mathcal{T}^{J\pm}_{\lambda'\lambda} =
  \frac{1}{2}\left[\mathcal{T}^{J}_{\lambda'\lambda} \pm
  \eta_1\eta_2(-1)^{s_1+s_2+\frac{1}{2}}
  \mathcal{T}^{J}_{\lambda'-\lambda}\right], 
\end{align}
where $\mathcal{T}^{J\pm}$ denotes the partial-wave transition
amplitude with total angular momentum $J$ and parity $(-1)^{J\pm 1/2}$.   
The prefactor $1/2$ ensures that no additional factor is required when
transforming back to the partial-wave component: 
\begin{align}
    \mathcal{T}^{J}_{\lambda'\lambda} =
  \mathcal{T}^{J+}_{\lambda'\lambda} +\mathcal{T}^{J-}_{\lambda'\lambda}.
\end{align}
At this stage, we want to emphasize that we do not need to decompose
the partial-wave component with definite parity in
Eq.~\eqref{eq:BS-1d}, as parity invariance is already imposed
in the effective Lagrangian and in the calculation of the
amplitudes shown in Eq.~\eqref{eq:ampi}. To study the 
dynamical generation of the resonances, we express the $\mathcal{T}$
matrix in the $IJL$ particle basis~\cite{Machleidt:1987hj}. The
relations between the $\mathcal{T}$ matrix elements in the two bases
are given by  
\begin{align}
  \mathcal{T}^{JS'S}_{L'L} = \frac{\sqrt{(2L+1)(2L'+1)}}{2J+1}
  \sum_{\lambda'_1\lambda'_2\lambda_1\lambda_2}
  \left(L'0S'\lambda'|J\lambda'\right)
  \left(s'_1\lambda'_1s'_2-\lambda'_2|S'\lambda'\right)
  \left(L0S\lambda|J\lambda\right)
  \left(s_1\lambda_1s_2-\lambda_2|S\lambda\right)
  \mathcal{T}^{J}_{\lambda'_1\lambda'_2,\lambda_1\lambda_2} .
\end{align}
In this work, we will only present the diagonal part
$\mathcal{T}^{JS}_{L}$ as it is relevant to particle production.   
\section{Results and discussions \label{sec:3}}
The molecular nature of the hidden charm pentaquarks is not a novel
concept. Prior to their discovery, numerous theoretical studies
predicted their existence as molecular states of the heavy meson and
heavy baryon
system~\cite{Wu:2010jy,Wang:2011rga,Wu:2012md,Xiao:2013yca}. The
hidden charm pentaquark states, $P_{c\bar{c}}$, discovered by the LHCb
Collaboration~\cite{LHCb:2015yax,LHCb:2019kea,LHCb:2021chn} are
positioned below various thresholds of the $\bar{D}\Sigma_c$,
$\bar{D}\Sigma_c^*$, and $\bar{D}^*\Sigma_c$ channels. Consequently,
many researchers considered the $P_{c\bar{c}}$ states to be molecular
states. Recently, however, the GlueX Collaboration~\cite{GlueX:2023pev} did
not observe any clear signal for the heavy pentaquarks in the $J/\psi
p$ invariant mass spectrum when measuring $J/\psi$ photoproduction off
the proton. Regarding this discrepancy, there is only one theoretical
work addressing it. Nakamura~\cite{Nakamura:2021qvy} proposed a reason
for the absence of the pentaquarks in the GlueX experiment: the hidden
charm pentaquarks, except for the $P_{c\bar{c}}$(4440), are cusp
structures arising from the kinematical effects of single and double
triangle diagrams. This can partially explain the disappearance of the
pentaquark peaks in $J/\psi$ photoproduction, although a pole diagram
is still needed to describe
$P_{c\bar{c}}$(4440)~\cite{Nakamura:2021qvy}. 
In contrast, we aim to elucidate why it is very difficult to observe
the signals of the pentaquarks from $J/\psi N$ photoproduction. We
will demonstrate that the origin of the discrepancy is dynamical, not
kinematical in this Section. 

Before presenting the numerical results, we first describe the fitting
procedure. While the coupling constants for various vertices are
theoretically fixed, the cutoff masses contain uncertainties due to
the lack of experimental data and theoretical estimation. As explained
in the previous Section, we fixed the cutoff masses using the relation
$\Lambda-m\approx(500-700)$ MeV. We will adjust these values
minimally. As shown in Table~\ref{tab:1}, we set $\Lambda-m =500$ MeV
for reactions involving heavy hadrons with low-lying masses, while 
choosing $\Lambda-m =700$ MeV for those with higher-lying masses. This
approach allows us to describe the four existing hidden charm
pentaquarks $P_{c\bar{c}}$ and predict three additional $P_{c\bar{c}}$
with larger masses.

As discussed in Ref.~\cite{Liu:2019tjn}, it is natural to expect that
there may be seven hidden charm pentaquark states in the $S$-wave with
negative parity, since we have seven different attractive channels:
$\bar{D}\Sigma_c(J^P=1/2^-)$, $\bar{D}\Sigma_c^*(J^P=3/2^-)$,
$\bar{D}^*\Sigma_c (J^P=1/2^-)$, $\bar{D}^*\Sigma_c (J^P=3/2^-)$,
$\bar{D}^*\Sigma_c^*(J^P=1/2^-)$, $\bar{D}^*\Sigma_c^*(J^P=3/2^-)$, and 
$\bar{D}^*\Sigma_c^*(J^P=5/2^-)$ with possible total angular
momenta considered. As will be shown below, we observe that
there are indeed seven peaks, among which six are identified as hidden
charm pentaquarks, while one peak exhibits a cusp structure rather
than a resonance. More interestingly, we find two additional $P$-wave
pentaquark resonances with positive parity. We will first investigate
the relevant transition amplitudes in the $S$-wave and examine the
nature of the hidden charm pentaquarks with negative parity. Then, we
will analyze the two pentaquark states with positive parity. Finally,
we will address the null results from the GlueX Collaboration. 
\subsection{Negative parity ($S$ wave interaction)}
We will first discuss the numerical results for the $P_{c\bar{c}}$'s
with negative parity. Though the parity for the $P_{c\bar{c}}$'s are
not yet experimentally given, the present results indicate that the
existing $P_{c\bar{c}}$'s must have the negative parity. 
To examine it, we define the partial-wave cross section with a given
spin-parity assignment as 
\begin{align}
    \sigma^{J\pm} = \mathrm{p}' \sum_{\lambda'\lambda} (2J+1)
  \left|\mathcal{T}_{\lambda'\lambda}^{J\pm}\right|^2 . 
\end{align}

\begin{figure}[htbp]
    \centering
    \includegraphics[scale=0.51]{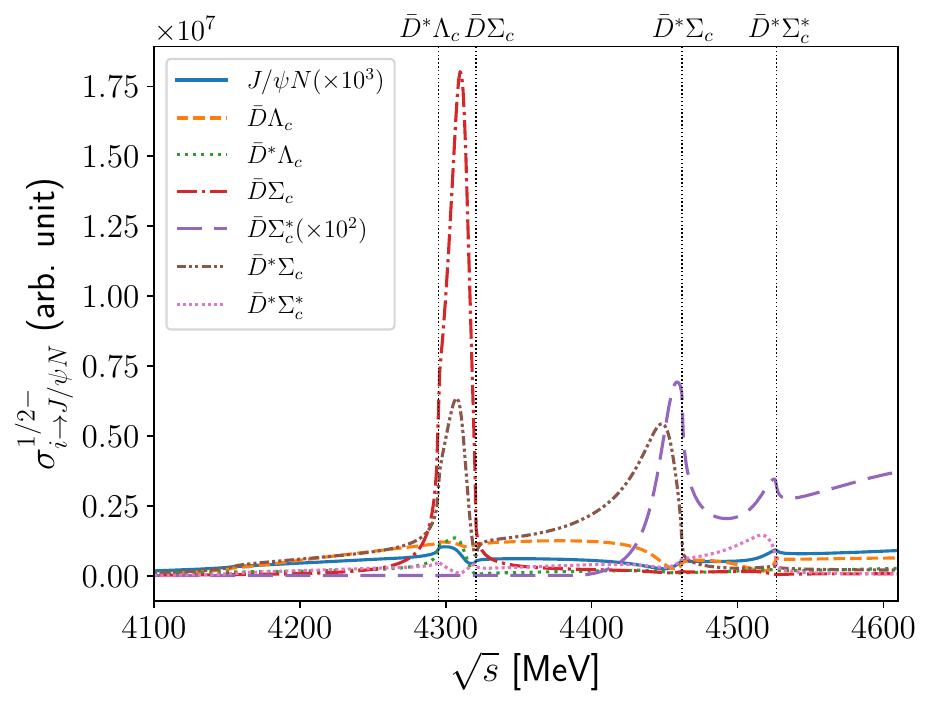}
    \includegraphics[scale=0.51]{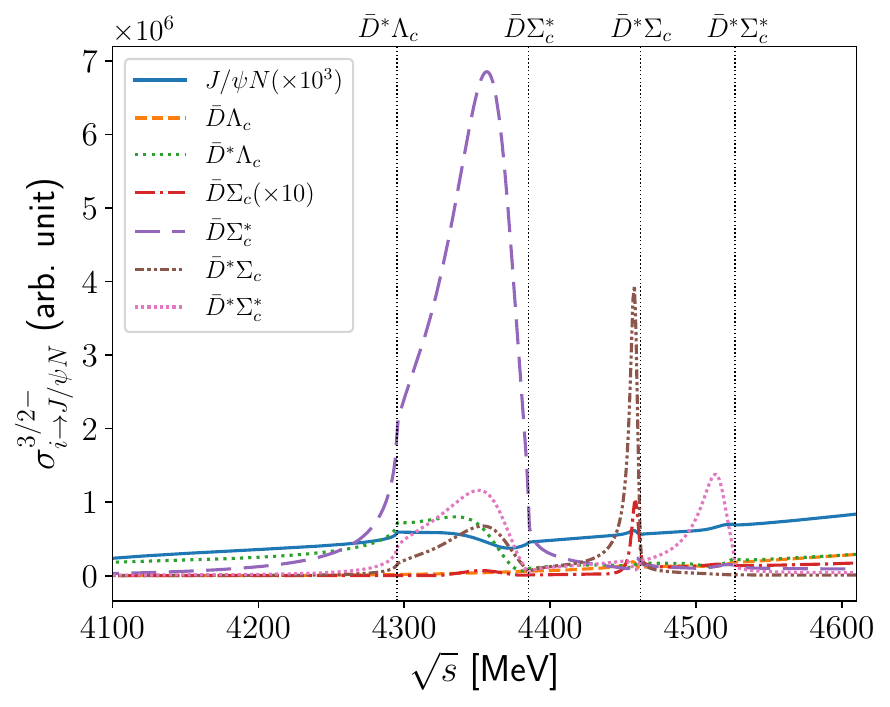}
    \includegraphics[scale=0.51]{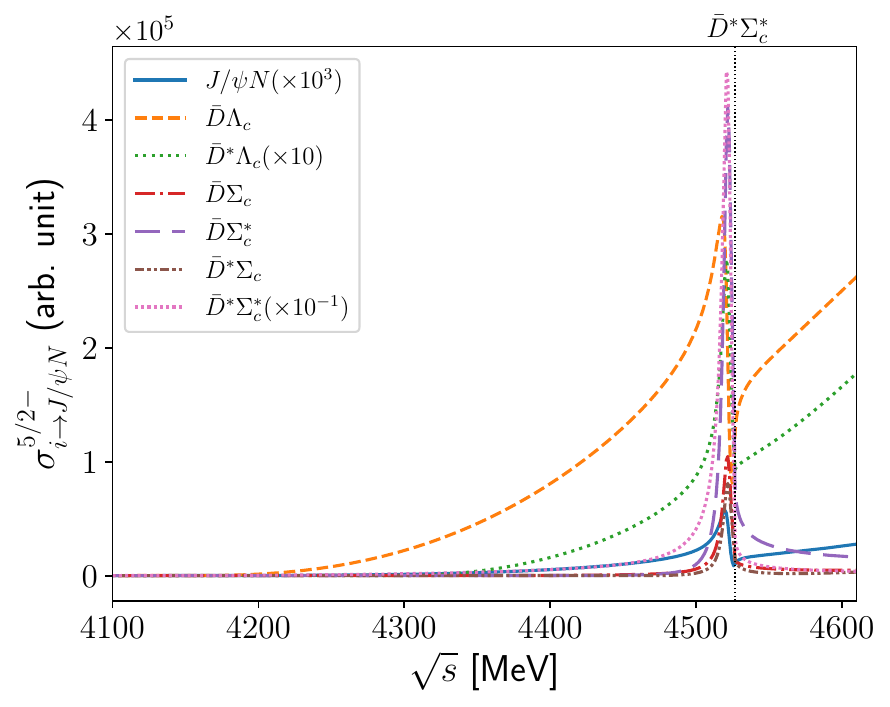}
    \caption{The partial-wave cross sections for the given total
      angular momenta $J=1/2,3/2,5/2$ with negative parity, which
      correspond to the spins and parities of $P_{c\bar{c}}$, as
      functions of total energy.}   
    \label{fig:3} 
  \end{figure}
In Fig.\ref{fig:3}, we present the partial-wave cross sections as
functions of energy in the CM frame, focusing on transitions from
several heavy meson and singly heavy baryon channels to the $J/\psi N$
channel. The upper left panel of Fig.\ref{fig:3} displays the
partial-wave transition cross sections with total angular momentum
$J=1/2$. The pentaquark state $P_{c\bar{c}}(4312)$ is clearly visible
in the $\bar{D}\Sigma_c\to J/\psi N$ transition. While the same
resonance appears in the $\bar{D}^*\Sigma_c\to J/\psi N$ transition,
its strength is weaker compared to the $\bar{D}\Sigma_c\to J/\psi N$
channel. No signal is observed in the $\bar{D}\Sigma_c^*\to J/\psi N$
channel, as only the $D$-wave contributes to this channel, which is too
weak to form $P_{c\bar{c}}(4312)$. Only vague hints of
$P_{c\bar{c}}(4312)$ are present in the $\bar{D} \Lambda_c$ and
$\bar{D}^* \Lambda_c$ channels. The $\bar{D}^*\Sigma_c^*\to J/\psi N$ 
transition exhibits only a destructive interference pattern. Notably,
there is almost no indication of $P_{c\bar{c}}(4312)$ in $J/\psi N$
scattering. Note that we have multiplied $\sigma_{J/\psi N}^{1/2}$ by
$10^3$. 
Below the $\bar{D}^*\Sigma_c$ threshold, we clearly observe a peak in
the $\bar{D}^*\Sigma_c \to J/\psi N$ transition, corresponding to
$P_{c\bar{c}}(4440)$. As shown by the dashed curve, a peak structure
is also found in the $\bar{D}\Sigma_c^* \to J/\psi N$ transition, but
it is multiplied by a factor of $10^2$. In other transitions, we
observe patterns of destructive interference. As with
$P_{c\bar{c}}(4312)$, we do not find any peak structure in $J/\psi N$
scattering. This observation may explain the null results for hidden
charm pentaquarks from the GlueX experiment, as will be discussed
later in detail. Interestingly, there are no clear signals for
pentaquark resonances below the $\bar{D}^*\Sigma_c^*$ threshold;
instead, we only observe cusp structures.

The upper right panel of Fig.~\ref{fig:3} presents the partial-wave
total cross sections for seven different transitions with
$J^P=3/2^-$. We observe clear peaks for $P_{c\bar{c}}(4380)$ in the
$\bar{D}\Sigma_c^*$, $\bar{D}^*\Sigma_c$, and $\bar{D}^*\Sigma_c^*$
channels. However, as in the previous case, we do not detect any peak
in $J/\psi N$ scattering. The $\bar{D} \Lambda_c$ and
$\bar{D}^*\Lambda_c$ channels do not exhibit any resonances. 
Below the $\bar{D}^*\Sigma_c$ threshold, we observe the
$P_{c\bar{c}}(4457)$ resonance in the $\bar{D}^*\Sigma_c$ channel. A
tiny resonance structure is also visible in the $\bar{D}\Sigma_c$
channel. Again, there is no peak structure in $J/\psi N$
scattering. Additionally, we identify a new resonance in the
$\bar{D}^*\Sigma_c^*\to J/\psi N$ transition, which has not yet been
observed experimentally. This new state could be designated as
$P_{c\bar{c}}(4517)$. In the lower panel of Fig.~\ref{fig:3}, we
observe $P_{c\bar{c}}(4522)(J^P=5/2^-)$, another state that has not been 
experimentally confirmed. Although the strength of the corresponding
peak appears weaker than the other $P_{c\bar{c}}$ resonances, it is
clearly visible in all channels. A unique feature of this new
resonance is its coupling to both $\bar{D}\Lambda_c$ and $\bar{D}^*
\Lambda_c$ channels.

\begin{figure}[htbp]
    \centering
    \includegraphics[scale=0.51]{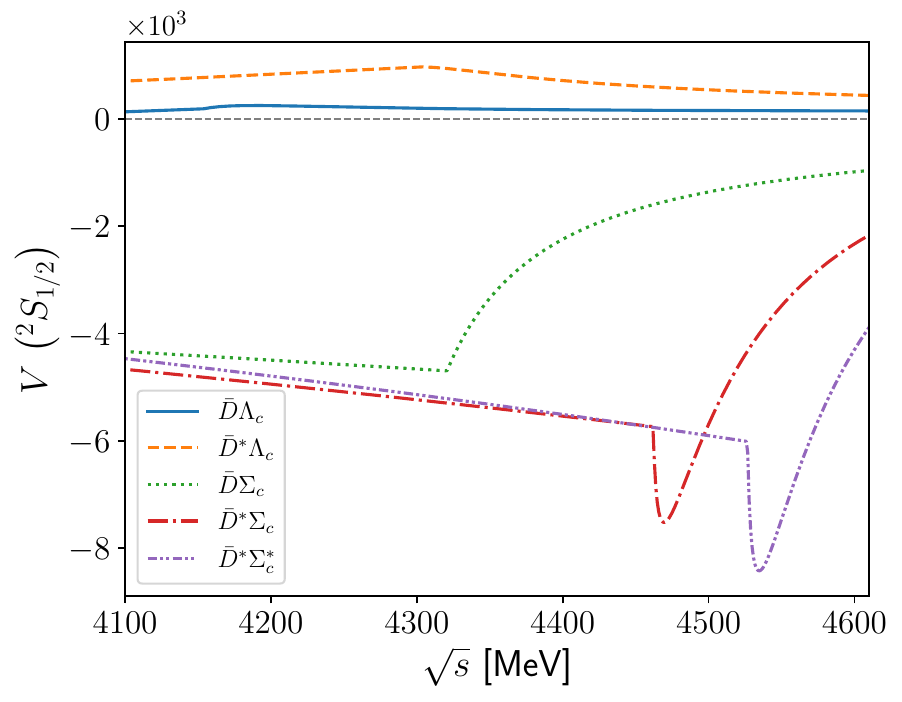}
    \includegraphics[scale=0.51]{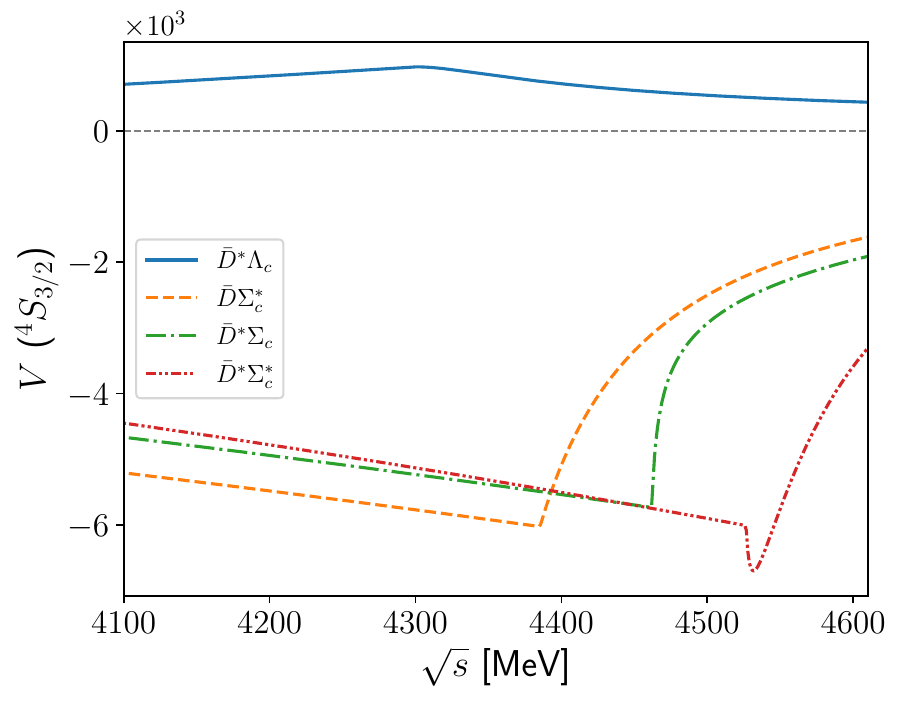}
    \includegraphics[scale=0.51]{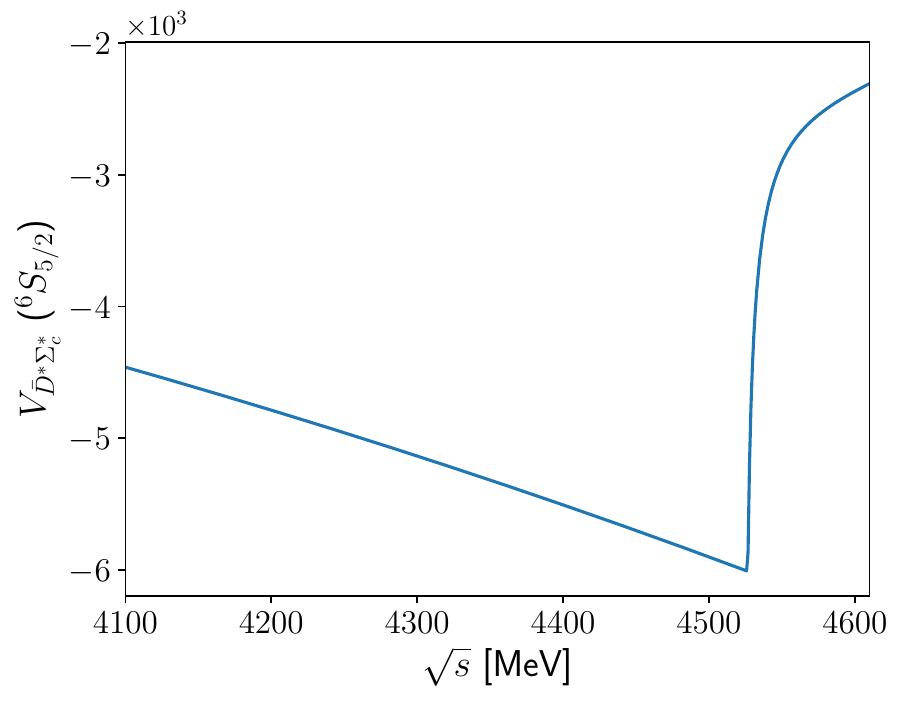}
    \caption{$S$-wave on-shell kernel amplitudes as functions of the
      total energy in the CM frame.}
    \label{fig:4}
  \end{figure}
In the partial-wave expansion, the $S$-wave provides the largest
contribution. This indicates that $S$-wave kernel amplitudes
generally serve as the primary source for the dynamical generation
of molecular states. 
Figure~\ref{fig:4} presents the results for the $S$-wave kernel
amplitudes in elastic scattering. The upper left, upper right, and
lower panels display the amplitudes for the ${}^2S_{1/2}$,
${}^4S_{3/2}$, and ${}^6S_{5/2}$ channels, respectively. 
We observe that the kernel amplitudes exhibit attractive interactions,
which are essential in producing the resonance structures shown in
Fig.~\ref{fig:3}. In contrast, the $\Lambda_c$, which belongs to the
baryon antitriplet, interacts repulsively with $\bar{D}$ and
$\bar{D}^*$. Consequently, resonances are not formed in the
$\bar{D}\Lambda_c$ and $\bar{D}^*\Lambda_c$ channels.

\begin{figure}[htbp]
    \centering
    \includegraphics[scale=0.51]{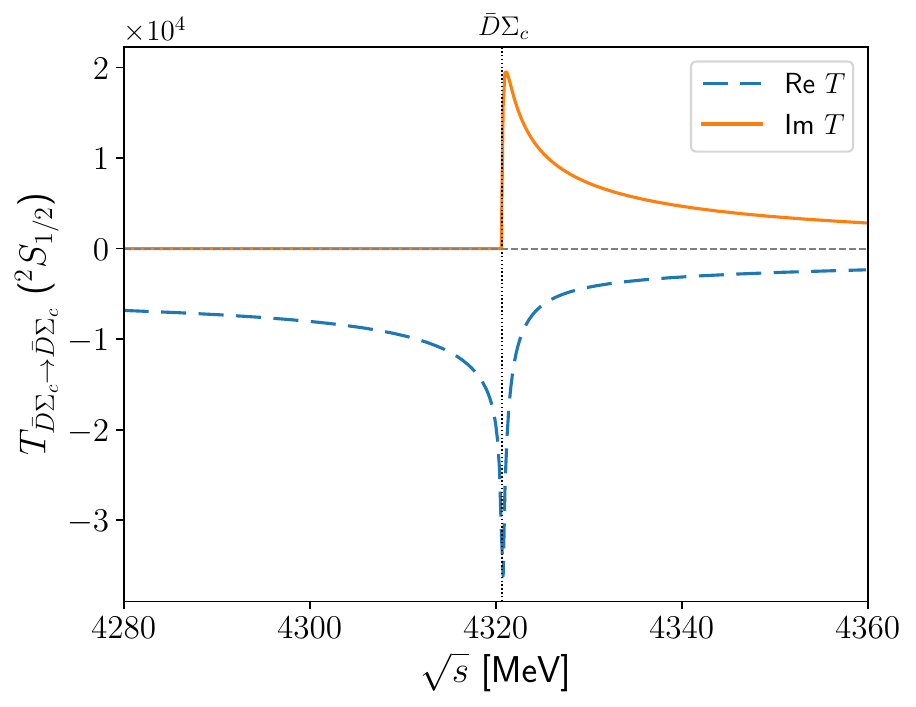}
    \includegraphics[scale=0.51]{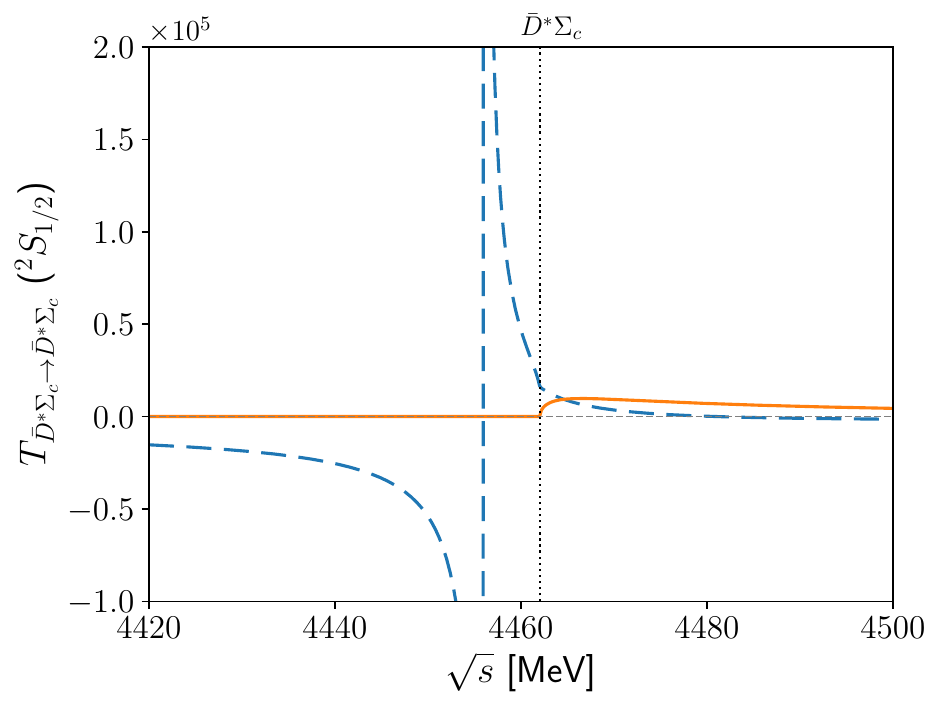}
    \includegraphics[scale=0.51]{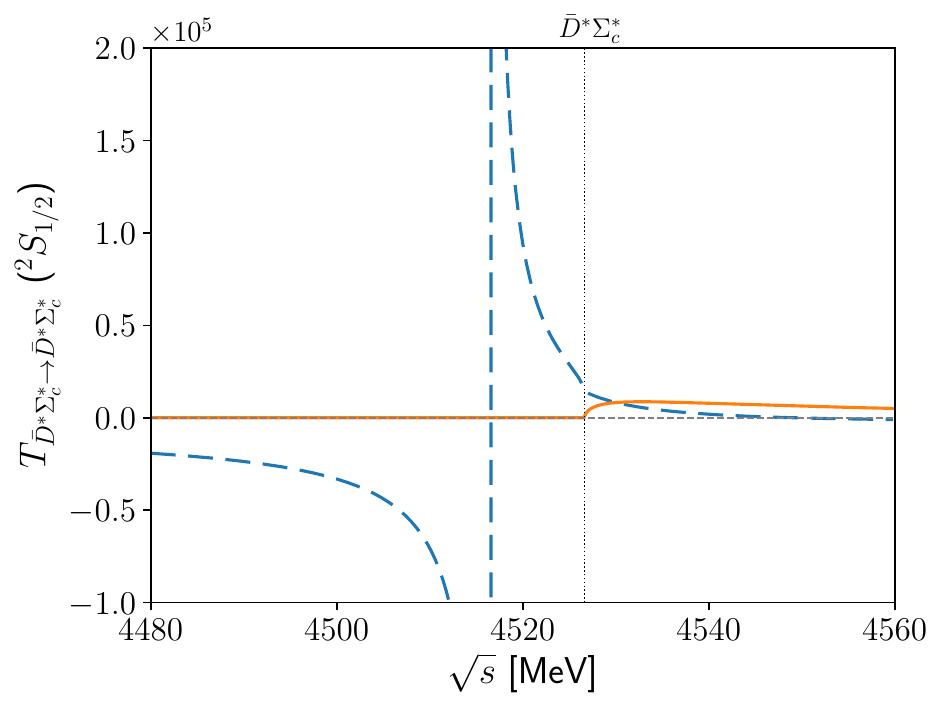}
    \includegraphics[scale=0.51]{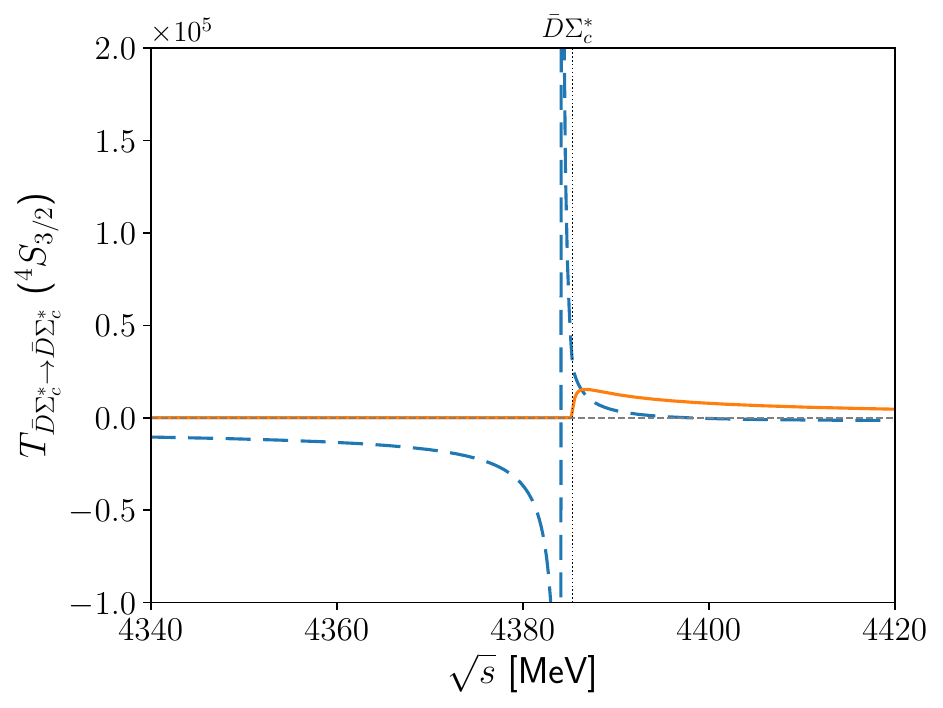}
    \includegraphics[scale=0.51]{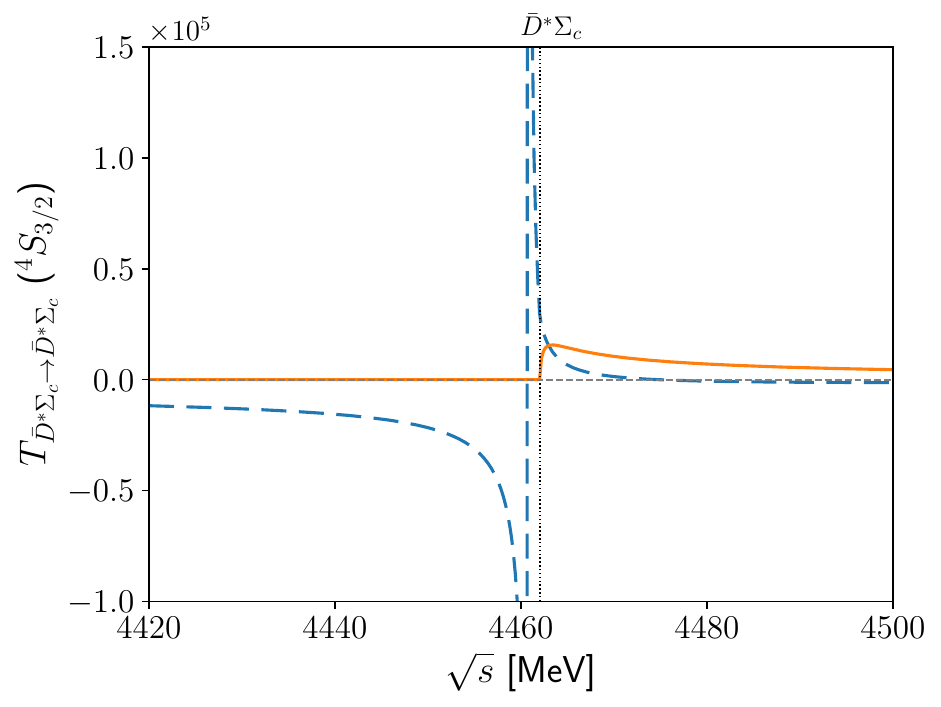}
    \includegraphics[scale=0.51]{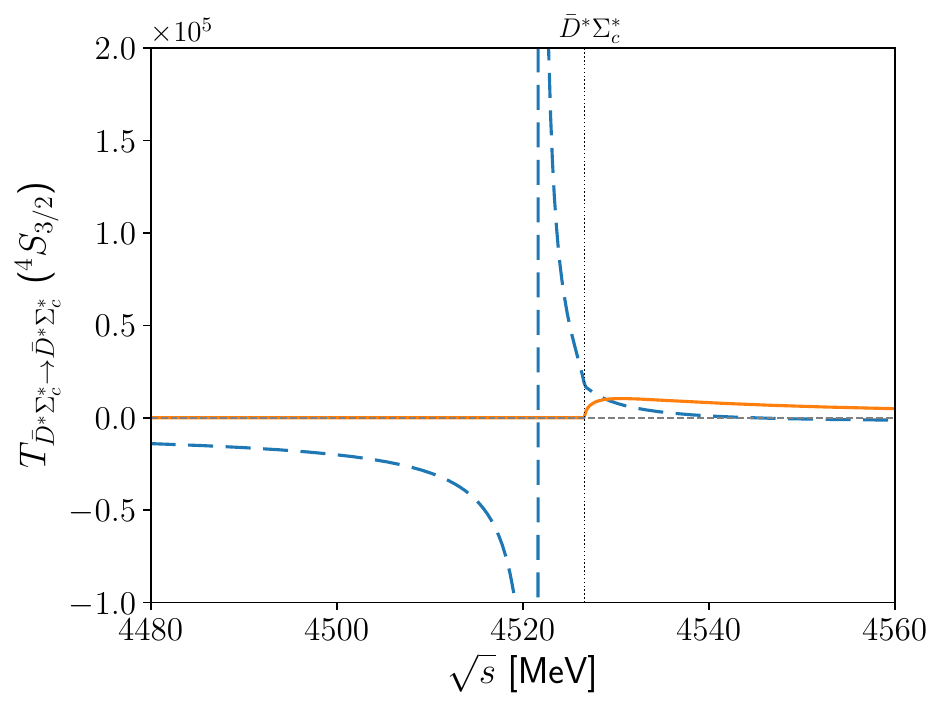}
    \includegraphics[scale=0.51]{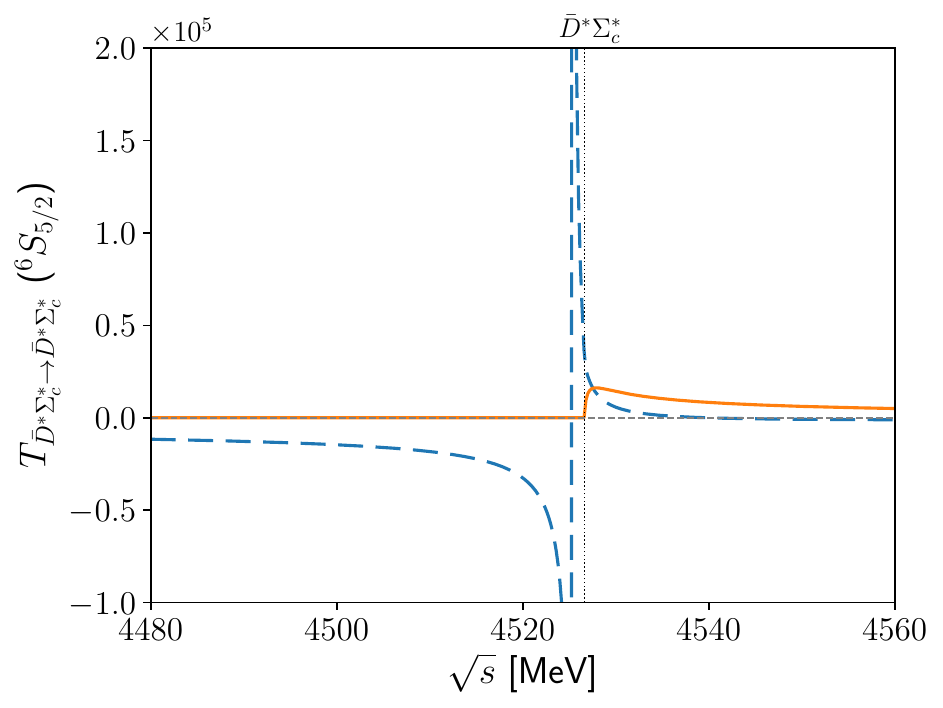}
    \caption{$S$-wave transition amplitudes generated by the single channel
      as functions of the total energy in the CM frame.} 
    \label{fig:5}
  \end{figure}
After examining the $S$-wave kernel amplitudes, we investigate the
corresponding transition amplitudes with the coupled-channel effects
turned off. This allows us to explicitly observe how the poles
corresponding to bound states appear. Figure~\ref{fig:5} presents the
results for the transition amplitudes across seven different
scattering channels. 
From the upper right panel to the lowest panel, we identify six bound
states in the following channels: $\bar{D}^*\Sigma_c(J=1/2)$,
$\bar{D}^*\Sigma_c^*(J=1/2)$, $\bar{D}\Sigma_c^*(J=3/2)$,
$\bar{D}^*\Sigma_c(J=3/2)$, $\bar{D}^*\Sigma_c^*(J=3/2)$, and
$\bar{D}^*\Sigma_c^*(J=5/2)$. 
We do not observe a bound state in the $\bar{D}\Sigma_c(J=1/2)$
channel, but we note an enhancement near the $\bar{D}\Sigma_c$
threshold. Although there is no bound state in this channel, a
resonance emerges when we include all possible coupled channels. This
observation emphasizes the significant role of coupled-channel effects
in the dynamical generation of hidden charm pentaquark states. We
observed a similar tendency in the dynamical generation of the $a_1$
meson~\cite{Clymton:2022jmv}.

\begin{table}[htbp]
  \caption{\label{tab:2} Hidden charm pentaquark states}
  \begin{ruledtabular}
  \centering\begin{tabular}{lcccccr}
   \multirow{2}{*}{$J^P$} & \multirow{2}{*}{Molecular states} & 
   \multicolumn{2}{c}{$\sqrt{s_R}=(M-i\Gamma/2)$ MeV} & \multicolumn{3}{c}{Known states} \\
   & & $M$ & $\Gamma$ & Name & $M$ & $\Gamma$
   \\\hline
     $1/2^{-}$ 
     & $[\bar{D}\Sigma_c]_{S=1/2}$ & $4312.71$ & $18.03$ & $P_{c\bar{c}}$(4312) & $4311.9^{+7.0}_{-0.9}$ & $10\pm 5$ \\
     & $[\bar{D}^*\Sigma_c]_{S=1/2}$ & $4457.75$ & $31.83$ & $P_{c\bar{c}}$(4440) & $4440^{+4}_{-5}$ & $21^{+10}_{-11}$ \\
     & $[\bar{D}^*\Sigma_c^*]_{S=1/2}$ & $-$ & $-$ & $-$ & $-$ & $-$ \\
     $3/2^{-}$ 
     & $[\bar{D}\Sigma_c^*]_{S=3/2}$ & $4369.15$ & $60.04$ & $P_{c\bar{c}}$(4380) & $4380\pm 30$ & $205\pm 90$ \\
     & $[\bar{D}^*\Sigma_c]_{S=3/2}$ & $4458.07$ & $6.28$ & $P_{c\bar{c}}$(4457) & $4457.3^{+4.0}_{-1.8}$ & $6.4^{+6}_{-2.8}$ \\
     & $[\bar{D}^*\Sigma_c^*]_{S=3/2}$ & $4516.90$ & $23.48$ & $-$ & $-$ & $-$ \\
     $5/2^{-}$ 
     & $[\bar{D}^*\Sigma_c^*]_{S=5/2}$ & $4522.05$ & $7.61$ & $-$ & $-$ & $-$ \\
  \end{tabular}
    \end{ruledtabular}
  \end{table}
The transition amplitudes obtained by solving the coupled integral
equations contain poles corresponding to the hidden charm pentaquark
states. By scanning these amplitudes in the complex energy plane, we 
can precisely locate the pole positions, which yield the masses and
widths of the pentaquarks. Table~\ref{tab:2} presents the masses and
widths of six hidden charm pentaquark states. Among these, four
resonances have been experimentally confirmed: $P_{c\bar{c}}(4312)$
and $P_{c\bar{c}}(4440)$ with $J=1/2$, and $P_{c\bar{c}}(4380)$ and
  $P_{c\bar{c}}(4457)$ with $J=3/2$. Thus, we predict
  the existence of $P_{c\bar{c}}(4517)$ and $P_{c\bar{c}}(4522)$.  

The cusp structure with total spin 1/2 below the $\bar{D}^*\Sigma_c^*$
threshold does not appear as a pole on the second Riemann
sheet. Further investigation of other sheets revealed its presence on
the upper sheet, formed by the branch point of the $\bar{D}^*\Sigma_c^*$
channel at $\sqrt{s_R}=(4529.27-i14.32)$ MeV. This behavior stems from
the coupled channel effect, which generates a repulsive interaction in
the molecular state. Consequently, the pole is pushed above the
$\bar{D}^*\Sigma_c^*$ channel threshold, becoming a virtual state. 
It is noteworthy that there exist more molecular states than those
observed experimentally, particularly in the region around 4.5
GeV. Two new states have been identified, corresponding to molecular
states of the $\bar{D}^*\Sigma_c^*$ system, while one is merely a
cusp. In contrast, Ref.\cite{Liu:2019tjn} predicted three new
states. With the exception of the cusp structure we found, the present
results align with those of Ref.\cite{Liu:2019tjn}. 

Table~\ref{tab:2} also compares the pole masses and widths from our
work with experimental data. While there is considerable agreement
overall, discrepancies emerge in some instances, such as for the
$P_{c\bar{c}}(4440)$ and $P_{c\bar{c}}(4380)$ resonances. This is not
unexpected, given that we did not fit the data. Furthermore, our
analysis reveals that the peak position and the real part of the pole
position are not identical. For instance, in the case of the
$\bar{D}^*\Sigma_c$ molecular state with $J=1/2$, the discrepancy is
as large as 14 MeV. These findings underscore the importance of
comprehensively examining the transition amplitudes to determine
resonance characteristics, rather than focusing solely on the peak
position, which can vary depending on the processes involved.

\begin{table}[htbp]
  \caption{\label{tab:3}Coupling strengths of the six $P_{c\bar{c}}$'s
    with $J^P=1/2^-$, $3/2^-$, and $5/2^-$.}  
  \begin{ruledtabular}
  \centering\begin{tabular}{lcccccc}
    $J^P$ & \multicolumn{2}{c}{$1/2^-$} & \multicolumn{3}{c}{$3/2^-$} & $5/2^-$ \\
    & $P_{c\bar{c}}(4312)$ & $P_{c\bar{c}}(4440)$ & $P_{c\bar{c}}(4380)$ & $P_{c\bar{c}}(4457)$ & $P_{c\bar{c}}(4517)$ & $P_{c\bar{c}}(4522)$ \\
    $\sqrt{s_R}$[MeV] & $4312.7-i9.0$ & $4457.7-i15.9$ & $4369.1-i30.0$ & $4458.1-i3.1$ & $4516.9-i11.7$ & $4522.1-i3.8$ 
   \\\hline
   $g_{J/\psi N({}^2S_J)}$            & $0.06+i0.04$   & $0.00+i0.10$  & $-$             & $-$           & $-$           & $-$           \\
   $g_{J/\psi N({}^2D_J)}$            & $-$            & $-$           & $0.02+i0.04$    & $0.00-i0.00$  & $0.00+i0.02$  & $0.01+i0.00$ \\
   $g_{J/\psi N({}^4S_J)}$            & $-$            & $-$           & $0.05+i0.03$    & $0.01+i0.00$  & $0.01+i0.00$  & $-$           \\
   $g_{J/\psi N({}^4D_J)}$            & $0.02+i0.02$   & $0.01+i0.03$  & $0.06+i0.09$    & $0.01-i0.01$  & $0.02-i0.03$  & $0.02+i0.00$ \\
   $g_{\bar{D}\Lambda_c({}^2S_J)}$    & $-0.24-i0.32$  & $-2.23-i1.96$ & $-$             & $-$           & $-$           & $-$           \\
   $g_{\bar{D}\Lambda_c({}^2D_J)}$    & $-$            & $-$           & $-0.07-i0.04$   & $0.48+i0.08$  & $0.43+i0.15$  & $0.68+i0.19$ \\
   $g_{\bar{D}^*\Lambda_c({}^2S_J)}$  & $7.19+i0.81$   & $3.69+i1.06$  & $-$             & $-$           & $-$           & $-$           \\
   $g_{\bar{D}^*\Lambda_c({}^2D_J)}$  & $-$            & $-$           & $0.09-i0.32$    & $-0.10+i0.03$ & $0.87+i0.09$  & $-0.31-i0.06$ \\
   $g_{\bar{D}^*\Lambda_c({}^4S_J)}$  & $-$            & $-$           & $11.51+i1.40$   & $-1.42-i0.45$ & $-3.16-i0.18$ & $-$           \\
   $g_{\bar{D}^*\Lambda_c({}^4D_J)}$  & $0.30-i0.34$   & $0.59+i0.09$  & $-0.16+i0.19$   & $0.30+i0.06$  & $0.34+i0.02$  & $0.50+i0.07$ \\
   $g_{\bar{D}\Sigma_c({}^2S_J)}$     & $-17.18-i2.76$ & $1.92-i1.63$  & $-$             & $-$           & $-$           & $-$           \\
   $g_{\bar{D}\Sigma_c({}^2D_J)}$     & $-$            & $-$           & $1.13+i0.62$    & $-1.10-i0.34$ & $-0.10+i0.10$ & $0.69+i0.22$ \\
   $g_{\bar{D}\Sigma_c^*({}^4S_J)}$   & $-$            & $-$           & $28.54+i8.76$   & $1.04-i0.43$  & $1.91-i0.60$  & $-$           \\
   $g_{\bar{D}\Sigma_c^*({}^4D_J)}$   & $-0.05-i0.02$  & $-1.59-i0.10$ & $0.01+i0.07$    & $1.13-i0.07$  & $-1.85-i0.69$ & $2.12+i0.38$ \\
   $g_{\bar{D}^*\Sigma_c({}^2S_J)}$   & $11.96+i4.72$  & $15.84+i9.45$ & $-$             & $-$           & $-$           & $-$           \\
   $g_{\bar{D}^*\Sigma_c({}^2D_J)}$   & $-$            & $-$           & $0.05+i0.09$    & $-0.00-i0.00$ & $0.98-i0.36$  & $-0.68+i0.06$ \\
   $g_{\bar{D}^*\Sigma_c({}^4S_J)}$   & $-$            & $-$           & $-9.01-i8.35$   & $11.66+i2.33$ & $-1.28+i0.33$ & $-$           \\
   $g_{\bar{D}^*\Sigma_c({}^4D_J)}$   & $-0.21-i0.10$  & $0.01-i0.03$  & $-0.05-i0.09$   & $0.00+i0.00$  & $0.20+i0.19$  & $0.98-i0.11$ \\
   $g_{\bar{D}^*\Sigma_c^*({}^2S_J)}$ & $-8.13-i2.12$  & $-4.79+i0.45$ & $-$             & $-$           & $-$           & $-$           \\
   $g_{\bar{D}^*\Sigma_c^*({}^2D_J)}$ & $-$            & $-$           & $-0.12-i0.15$   & $-0.00+i0.00$ & $-0.01-i0.02$ & $-0.00-i0.00$ \\
   $g_{\bar{D}^*\Sigma_c^*({}^4S_J)}$ & $-$            & $-$           & $-17.07-i15.28$ & $0.85+i1.79$  & $16.69+i5.72$ & $-$           \\
   $g_{\bar{D}^*\Sigma_c^*({}^4D_J)}$ & $-0.25-i0.09$  & $-0.04-i0.03$ & $0.16+i0.23$    & $-0.01-i0.01$ & $-0.01-i0.02$ & $0.00+i0.00$ \\
   $g_{\bar{D}^*\Sigma_c^*({}^6S_J)}$ & $-$            & $-$           & $-$             & $-$           & $-$           & $12.24+i3.02$ \\
   $g_{\bar{D}^*\Sigma_c^*({}^6D_J)}$ & $0.08+i0.01$   & $0.03-i0.04$  & $0.06+i0.07$    & $0.00+i0.00$  & $0.00+i0.01$  & $0.00+i0.01$ \\
  \end{tabular}
    \end{ruledtabular}
  \end{table}
Table \ref{tab:3} lists the numerical results for the coupling
strengths of the six $P_{c\bar{c}}$ states with negative parity,
demonstrating the intensity of their couplings to all possible decay
channels. The $P_{c\bar{c}}(4312)$ state, for instance, couples most
strongly to the $\bar{D}\Sigma_c$ channel. This suggests that
$P_{c\bar{c}}(4312)$ is likely a molecular state composed of $\bar{D}$
and $\Sigma_c$, given that its mass is below the $\bar{D}\Sigma_c$
threshold. Notably, it also exhibits strong coupling to the $S$-wave
$\bar{D}^*\Sigma_c({}^2S_{1/2})$ and
$\bar{D}^*\Sigma_c^*({}^2S_{1/2})$ channels. The coupling to
$\bar{D}^*\Lambda_c({}^2S_{1/2})$ is 
particularly interesting, as it indicates that $P_{c\bar{c}}(4312)$
contains a mixture of $\bar{D}^*\Lambda_c({}^2S_{1/2})$ and
$\bar{D}\Sigma_c({}^2S_{1/2})$. Furthermore, the
$\bar{D}^*\Lambda_c({}^2S_{1/2})$ channel contributes to the decay of
$P_{c\bar{c}}(4312)$. While experimental observations have shown that
$P_{c\bar{c}}(4312)$ decays into $J/\psi$ and $N$, these results
suggest that its decay into $\bar{D}^*$ and $\Lambda_c$ should be even
more pronounced. 

The $P_{c\bar{c}}(4440)$ exhibits strong coupling to the $S$-wave
$\bar{D}^* \Sigma_c({}^2S_{1/2})$ channel, suggesting it is a
molecular state of $\bar{D}^*$ and $\Sigma_c$. It also couples to the
$S$-wave $\bar{D}^* \Sigma_c^*({}^2S_{1/2})$, $\bar{D}^* \Lambda_c
({}^2S_{1/2})$, and $\bar{D} \Lambda_c$ channels, indicating a mixed
state. Consequently, $P_{c\bar{c}}(4440)$ can kinematically decay into
$\bar{D}$ and $\Lambda_c$, as well as $\bar{D}^*$ and $\Lambda_c$. 
The $P_{c\bar{c}}(4380)$ resonance couples most strongly to the
$\bar{D}\Sigma_c^*({}^4S_{3/2})$ channel, with significant
contributions from $\bar{D}^* \Sigma_c^*({}^4S_{3/2})$ and
$\bar{D}^*\Lambda_c({}^4S_{3/2})$ channels. This suggests
$P_{c\bar{c}}(4380)$ is a $\bar{D}\Sigma_c^*$ molecular state mixed
with $\bar{D}^* \Sigma_c^*$ and $\bar{D}^*\Lambda_c$ components. 
The $P_{c\bar{c}}(4457)$ resonance is predominantly governed by the
$S$-wave $\bar{D}^*\Sigma_c({}^4S_{3/2})$ channel, with minor
contributions from $\bar{D}^*\Lambda_c({}^4S_{3/2})$,
$\bar{D}\Sigma_c({}^2D_{3/2})$, $\bar{D}\Sigma_c^*({}^4S_{3/2})$, and
$\bar{D}\Sigma_c^*({}^4D_{3/2})$. This indicates that
$P_{c\bar{c}}(4457)$ can be considered a $\bar{D}^*\Sigma_c$ molecular
state. The last two columns of Table~\ref{tab:3} list the numerical results
for the coupling strengths of the new resonant states
$P_{c\bar{c}}(4517, J=3/2)$ and $P_{c\bar{c}}(4522, J=5/2)$. These
results suggest that these two hidden charm pentaquarks are likely
$S$-wave $\bar{D}^*\Sigma_c^*$ molecular states. 

\subsection{Positive parity}
A great virtue of the present coupled-channel formalism is that we
can also predict $P$-wave pentaquark states with positive parity.
While the $S$-wave contribution is the most dominant one, the $P$-wave
interaction is also strong enough to form a resonance. For example,
the first baryonic resonance $\Delta$ isobar is also a $P$-wave
resonance from $\pi N$ scattering. The nature of these pentaquarks is
distinguished from those with negative parity. The $P$-wave hidden
charm pentaquark states with positive parity emerge from the
constructive interference of various channels, which will be discussed
below.  

\begin{figure}[htbp]
    \centering
    \includegraphics[scale=0.51]{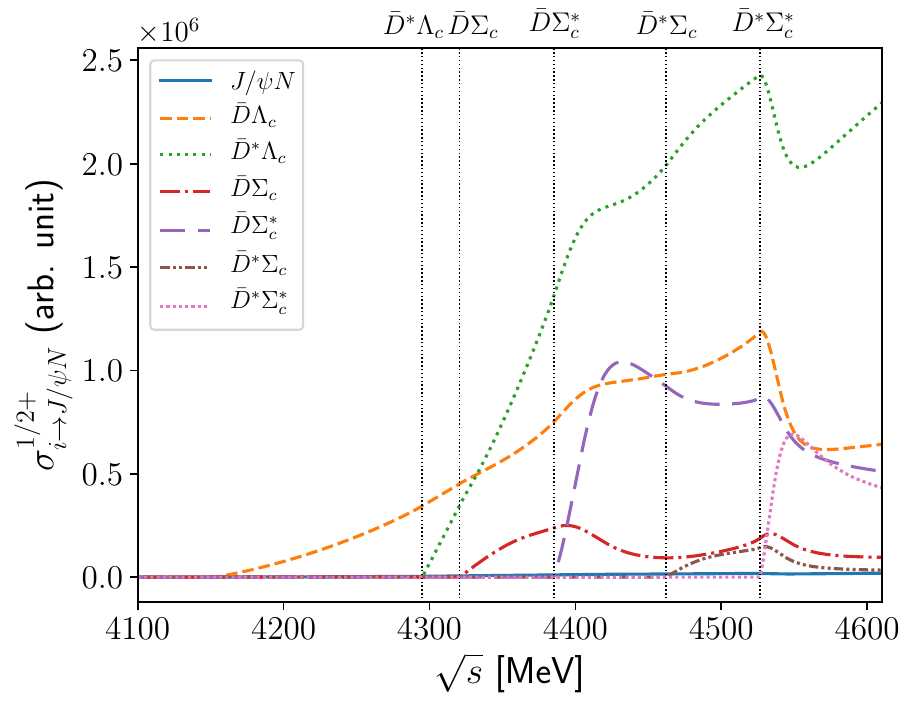}
    \includegraphics[scale=0.51]{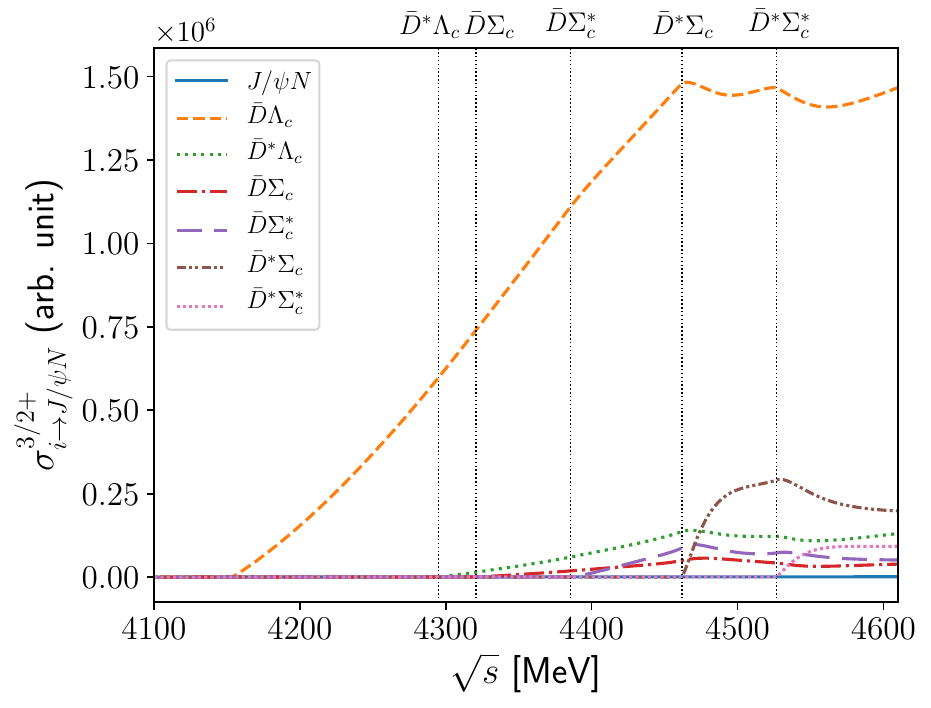}
    \includegraphics[scale=0.51]{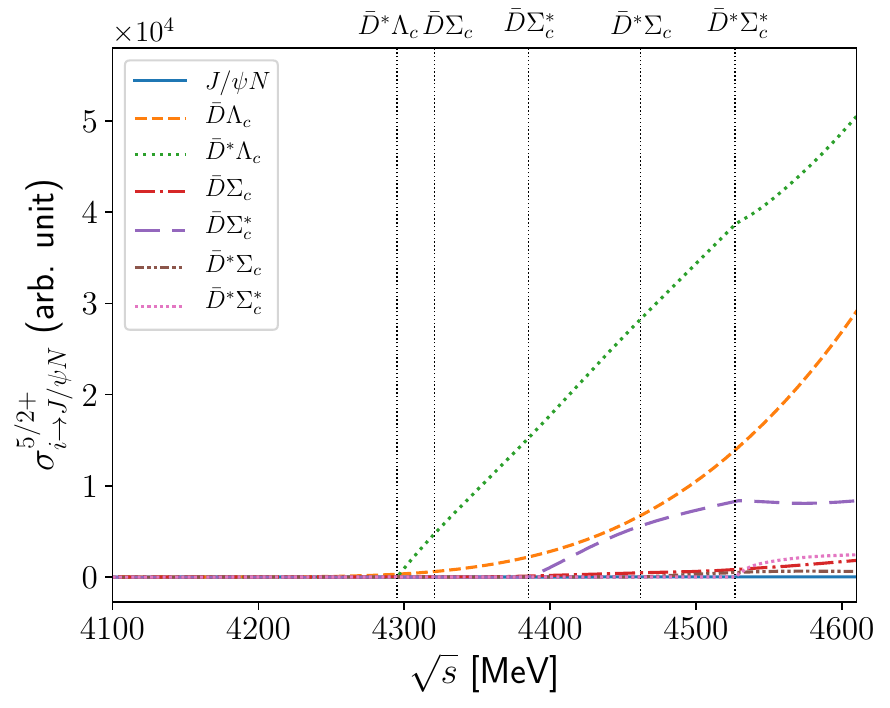}
    \caption{The partial-wave total cross sections for the total
      angular momenta $J=1/2$, $3/2$, and $5/2$ with positive parity
      as functions of the total energy in the CM frame.}  
    \label{fig:6}
  \end{figure}
The upper left panel of Fig.\ref{fig:6} illustrates the partial-wave
total cross sections for transitions from a heavy meson and a singly
heavy baryon to $J/\psi N$ with total spin $J=1/2$. The complex energy
plane reveals poles at $(4401.11-i35.21)$ MeV and $(4532.57-i17.10)$
MeV, corresponding to the resonances shown in this panel. These
resonances exhibit markedly different characteristics compared to
those with total spin $J=1/2$ and negative parity, depicted in the
upper left panel of Fig.\ref{fig:3}. While Fig.~\ref{fig:3} clearly
shows $P_{c\bar{c}}(4312)$ and $P_{c\bar{c}}(4440)$ as molecular
states, these two positive-parity resonances cannot be identified
within a single molecular picture. The upper right panel of
Fig.~\ref{fig:6} displays the partial-wave total cross sections for
$J^P=3/2^+$. Two peak structures appear near the $\bar{D}^*\Sigma_c$
and $\bar{D}^*\Sigma_c^*$ thresholds. However, the absence of
corresponding poles in the second Riemann sheet indicates no
resonances for $J^P=3/2^+$. The lower panel of Fig.~\ref{fig:6} shows
no peak structure for $J^P=5/2^+$. 

\begin{table}[htbp]
  \caption{\label{tab:4}Coupling strengths of $P_{c\bar{c}}$'s with $J^P=1/2^+$.}
  \begin{ruledtabular}
  \centering\begin{tabular}{lrr}
    $J^P$ & \multicolumn{2}{c}{$1/2^+$} \\
    $\sqrt{s_R}$[MeV] & $4401.1-i35.2$ & $4532.6-i17.1$  
   \\\hline
   $g_{J/\psi N({}^2P_J)}$            & $0.00+i0.00$  & $0.05+i0.02$ \\
   $g_{J/\psi N({}^4P_J)}$            & $0.15-i0.08$  & $0.09+i0.03$ \\
   $g_{\bar{D}\Lambda_c({}^2P_J)}$    & $-0.20+i0.59$ & $0.95-i0.69$ \\
   $g_{\bar{D}^*\Lambda_c({}^2P_J)}$  & $0.84-i0.73$  & $1.22-i0.89$ \\
   $g_{\bar{D}^*\Lambda_c({}^4P_J)}$  & $1.53-i1.92$  & $0.78-i0.48$ \\
   $g_{\bar{D}\Sigma_c({}^2P_J)}$     & $-1.94+i0.56$ & $-0.78-i0.35$ \\
   $g_{\bar{D}\Sigma_c^*({}^4P_J)}$   & $-$           & $0.78+i0.37$ \\
   $g_{\bar{D}^*\Sigma_c({}^2P_J)}$   & $-$           & $1.08-i0.10$ \\
   $g_{\bar{D}^*\Sigma_c({}^4P_J)}$   & $-$           & $0.23-i0.99$ \\
   $g_{\bar{D}^*\Sigma_c^*({}^2P_J)}$ & $-$           & $-2.50+i5.15$ \\
   $g_{\bar{D}^*\Sigma_c^*({}^4P_J)}$ & $-$           & $-1.96+i2.46$ \\
   $g_{\bar{D}^*\Sigma_c^*({}^6P_J)}$ & $-$           & $-$ \\
   $g_{\bar{D}^*\Sigma_c^*({}^6F_J)}$ & $-$           & $0.27-i0.16$ \\
  \end{tabular}
    \end{ruledtabular}
  \end{table}
Table~\ref{tab:4} lists the coupling strengths of the two pentaquark
resonances with $J^P=1/2^+$ to all possible transition channels. The
resonance at $(4401.11-i35.21)$ MeV primarily couples to
$\bar{D}\Sigma_c({}^2P_{1/2})$, $\bar{D}^*\Lambda_c({}^4P_{1/2})$, and
$\bar{D}^*\Lambda_c({}^2P_{1/2})$, while showing weak coupling to other
channels. Notably, this resonance does not couple to transition
channels with higher energy than its mass, distinguishing it from the
negative-parity $P_{c\bar{c}}$ states. The resonance at
$(4532.57-i17.10)$ MeV, detailed in the last column of
Table~\ref{tab:4}, exhibits a more complex nature. Its formation 
involves eight different transition channels, indicating a intricate
structure. These two resonances lack a clear molecular structure,
suggesting they may be candidates for a \emph{genuine} pentaquark
configuration. A more comprehensive analysis of these states will be
presented in future work.
\subsection{Null results from the GlueX experiment}
\begin{figure}[htbp]
    \centering
    \includegraphics[scale=0.51]{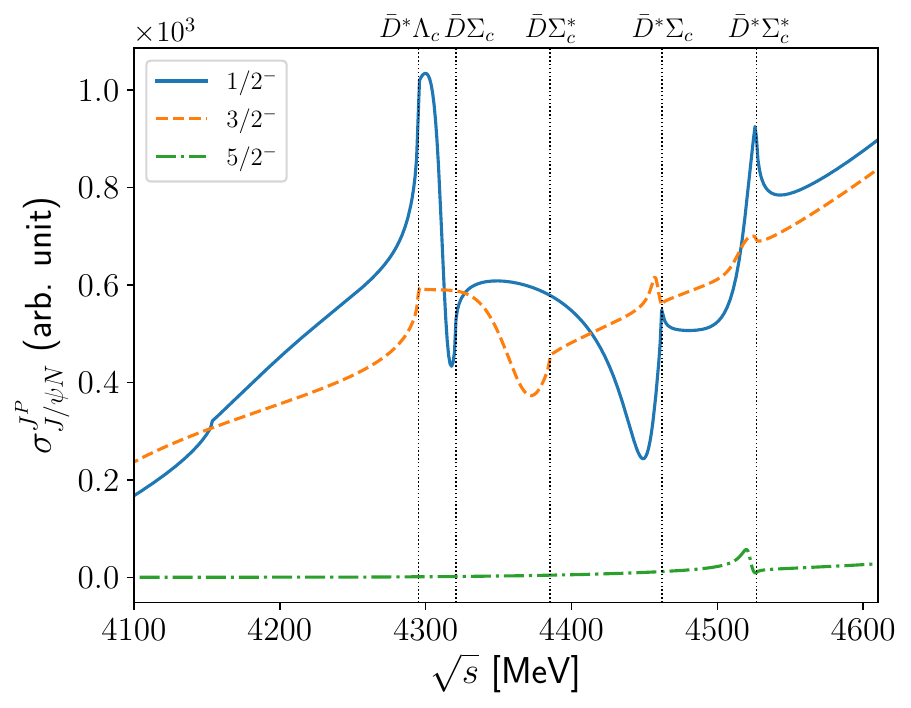}
    \includegraphics[scale=0.51]{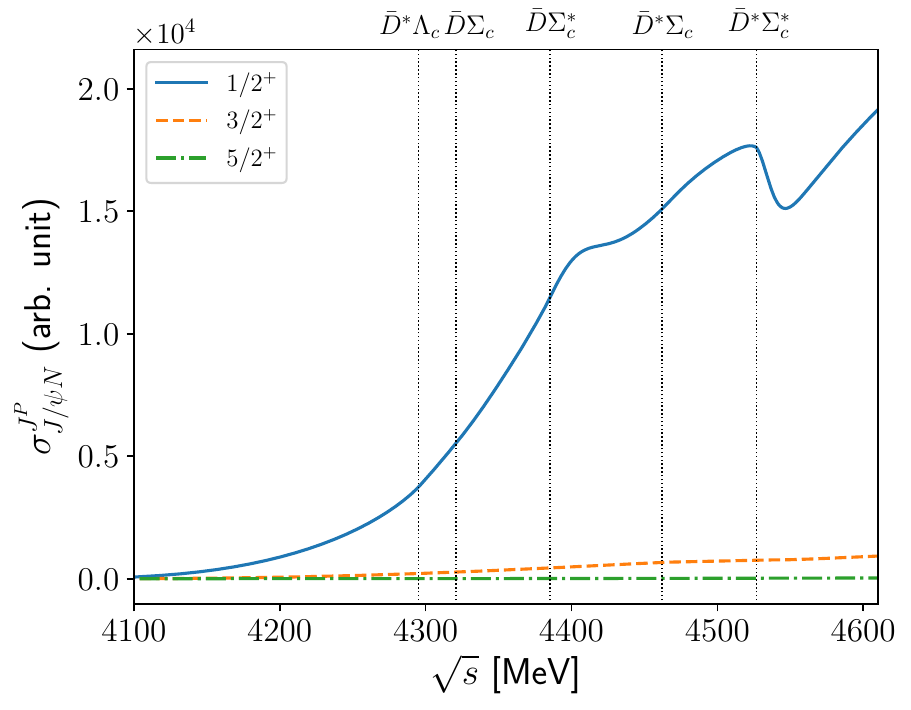}
    \caption{
Partial-wave total cross sections for $J/\psi N$ scattering with
$J=1/2$, $3/2$, and $5/2$. Left panel: $\sigma_{J/\psi N}^{J^-}$ for
negative parity states. Right panel: $\sigma_{J/\psi N}^{J^+}$ for
positive parity states. Both are plotted as functions of the total
energy in the CM frame. 
}
\label{fig:7}
\end{figure}
\begin{figure}[htbp]
    \centering
    \includegraphics[scale=0.51]{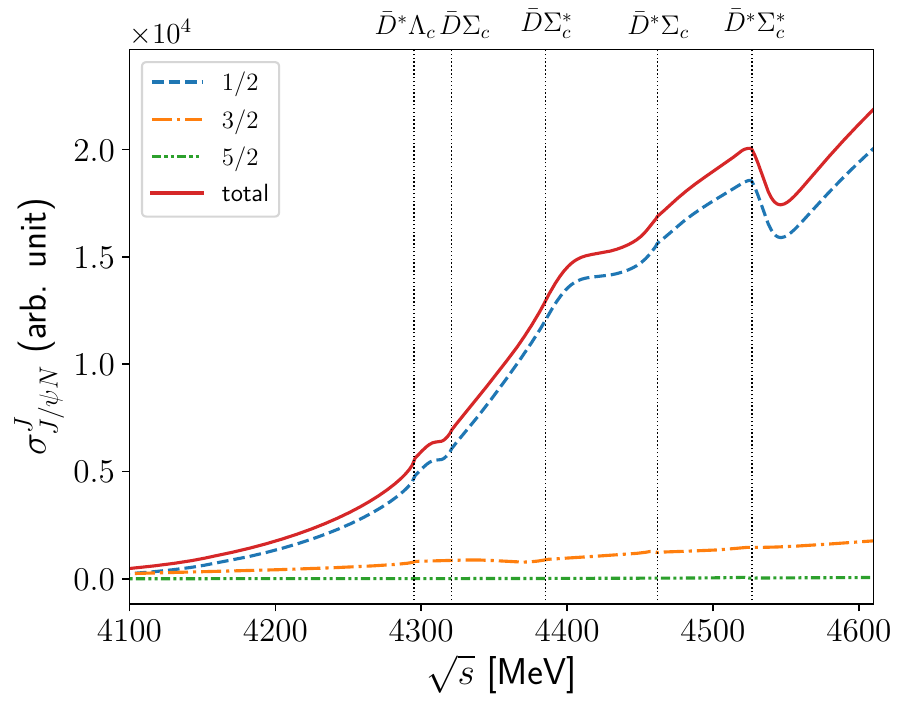}
    \caption{
Partial-wave total cross sections for $J/\psi N$ scattering with
$J=1/2$, $3/2$, and $5/2$ as functions of the total
energy in the CM frame. 
} 
    \label{fig:8}
\end{figure}

The GlueX Collaboration recently reported null results for
$P_{c\bar{c}}$ states in $J/\psi$ photoproduction off the
proton~\cite{GlueX:2023pev}. Our results may explain this absence. The
key factor is the transition amplitude for $J/\psi N$ scattering,
which resembles $J/\psi N$ photoproduction in the vector meson
dominance picture. The rescattering equation for $J/\psi N$
photoproduction can be expressed as: 
\begin{align}
T_{\gamma N \to J\psi N} = V_{\gamma N \to J\psi N} +
\sum_{j}V_{\gamma N \to j}G_j \mathcal{T}_{j\to J\psi N},
\label{eq:53}
\end{align}
where $j$ represents the seven channels associated with $P_{c\bar{c}}$
production. In Eq.~\eqref{eq:53}, $V_{\gamma N\to J/\psi N}$ is likely
the dominant kernel, as the photon is strongly coupled to the $J/\psi$
because of its vector nature ($J^{PC}=1^{--}$). 
Figure~\ref{fig:7} illustrates the partial-wave total cross sections
for $J/\psi N$ scattering with spin $J=1/2$, $3/2$, and $5/2$ for
negative (left panel) and positive (right panel) parities. The peaks
corresponding to negative-parity $P_{c\bar{c}}$ states appear as dip
structures. However, this alone does not explain the GlueX results, as
peaks remain visible, particularly for $P_{c\bar{c}}(4312)$. 
The positive-parity results show bump structures approximately ten
times larger than those with negative parity. Figure~\ref{fig:8}
presents the combined partial-wave total cross sections.
Consequently, $P_{c\bar{c}}(4312)$ state smears out along with all
other $P_{c\bar{c}}$'s with negative parity, as illustrated in
Fig.~\ref{fig:8}. 

\begin{figure}[htbp]
    \centering
    \includegraphics[scale=0.51]{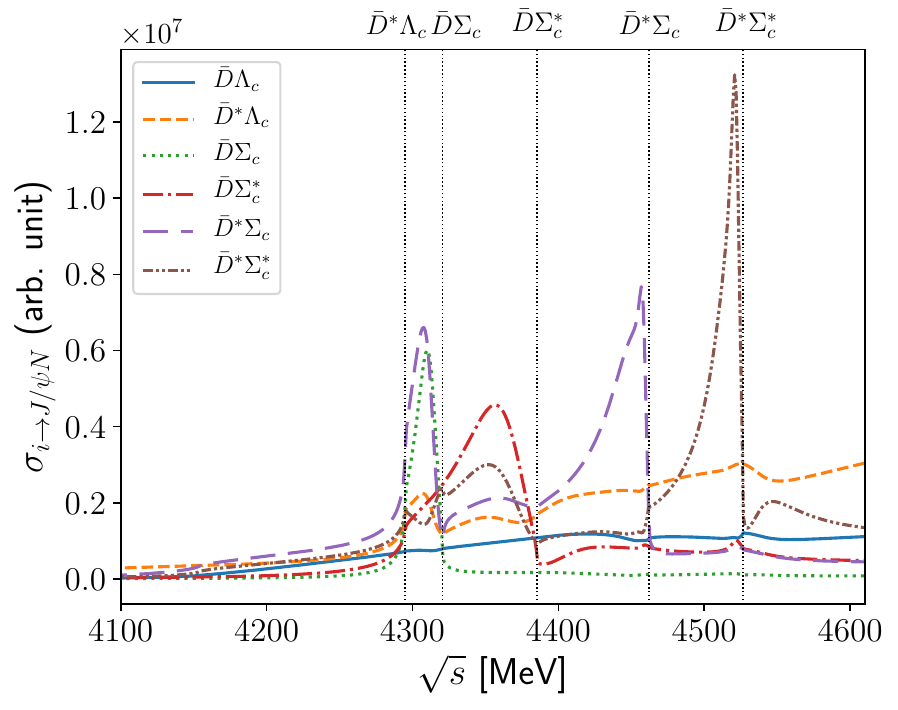}
    \includegraphics[scale=0.51]{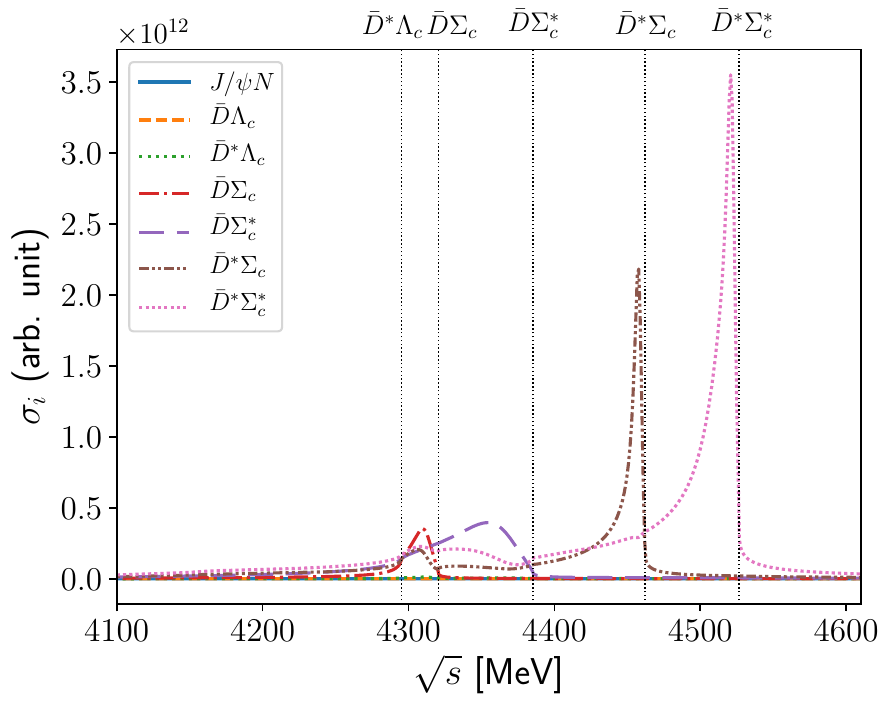}
    \caption{Total cross sections for the seven different
      transitions as functions of the total energy in the CM
      frame. Left panel: total cross sections for the transitions from
      a heavy meson and a singly heavy baryon to $J/\psi N$. Right
      panel: total cross sections for eleastic scattering of seven
      different channels.} 
    \label{fig:9}
  \end{figure}
We can explain both the absence of $P_{c\bar{c}}$ states in the GlueX
experiment and their observation in the LHCb experiment. In
$\Lambda_b\to J/\psi pK^-$ decay, six transition channels are
unsuppressed, differing only in kinematic factors. The left panel of
Fig.\ref{fig:9} shows the total cross sections for these six
transition channels, clearly displaying peak structures corresponding
to hidden charm pentaquark states. As previously discussed,
$P_{c\bar{c}}(4312,J^P=1/2^-)$ is clearly observed below the
$\bar{D}\Sigma_c$ threshold in the $\bar{D}\Sigma_c\to J/\psi N$ and
$\bar{D}^* \Sigma_c\to  J/\psi N$ transitions. It is also evident in
the $\bar{D}^*\Lambda_c \to J/\psi N$ transition, though the magnitude
of the resonance is smaller than in the two aforementioned
transitions. Below the $\bar{D}\Sigma_c^*$ threshold, we find the
$P_{c\bar{c}}(4380, J^P=3/2^-)$ resonance in the $\bar{D}\Sigma_c^*\to
J/\psi N$ and $\bar{D}^*\Sigma_c^*\to J/\psi N$ transitions. The
$P_{c\bar{c}}(4440, J^P=1/2^-)$ resonance appears clearly below the
$\bar{D}^*\Sigma_c$ 
threshold. The $P_{c\bar{c}}(4457,J^P=3/2^-)$ is located
just below the $\bar{D}^*\Sigma_c$ threshold, overlapping with
the $P_{c\bar{c}}(4440, J^P=1/2^-)$ state. Additionally, two new hidden
charm pentaquark states, which are closely spaced, are
shown below the $\bar{D}^*\Sigma_c^*$ threshold in the
$\bar{D}^*\Sigma_c^* \to J/\psi N$ transition. Since the magnitudes of
the predicted two pentaquark states with positive parity are notably
smaller than those with negative parity, we do not see them in the
total transition cross sections. The right panel of Fig.\ref{fig:9}
depicts the total cross sections for seven elastic scattering
channels, revealing only negative-parity pentaquark states, as in the
case of the transitions. While the total cross section for $J/\psi N$
scattering is very small in comparison with other channels, we want to
mention that the predicted two pentaquark states with positive parity
can be also seen in it, as shown in Fig.~\ref{fig:8}.

\section{Summary and conclusions\label{sec:4}}
In this work, we investigated hidden-charm pentaquark states using an
off-shell coupled-channel formalism involving heavy meson and singly
heavy baryon scattering. Our analysis identified seven distinct peaks
related to molecular states of heavy mesons $\bar{D}$ ($\bar{D}^*$) and
singly heavy baryons $\Sigma_c$ ($\Sigma_c^*$). Among these, six are
identified as resonances, while one exhibits a cusp structure. Four of
these peaks can be associated with known $P_{c\bar{c}}$ states:
$P_{c\bar{c}}(4312)$, $P_{c\bar{c}}(4380)$, $P_{c\bar{c}}(4440)$, and
$P_{c\bar{c}}(4457)$. Additionally, we predicted two new resonances
with masses around 4.5 GeV, which we interpret as $\overline{D}^*
\Sigma_c^*$ molecular states. 
Our study revealed that these pentaquark states undergo significant
modifications in the $J/\psi N$ elastic channel, with some even
disappearing due to interference from the positive parity channel. 
The combined partial-wave total cross sections for $J/\psi N$
scattering demonstrate how the $P_{c\bar{c}}(4312)$ state, along with
other negative parity $P_{c\bar{c}}$ states, are smeared out due to
interference with positive parity contributions. This contrasts with
the clear visibility of pentaquark states in transitions from heavy
meson and singly heavy baryon channels to $J/\psi N$. These findings
provide potential insight into the absence of pentaquark states in
$J/\psi$ photoproduction observed by the GlueX collaboration, while
also explaining their observation in LHCb experiments. 

We also identified two $P$-wave pentaquark states with positive
parity, which may be candidates for a genuine pentaquark
configuration. However, several important points require further
investigation. Further theoretical investigations are required to
fully explain the disappearance of the $P_{c\bar{c}}$ states in
photoproduction, which will be the subject of our next project. It may
be possible to observe the LHCb $P_{c\bar{c}}$ states in the open
charm final state of photoproduction, although this presents
significant experimental challenges. Furthermore, the emergence of the
$P_{c\bar{c}}(4330)$ state in the $B_s^0\to J/\psi p\bar{p}$ decay
channel cannot be explained within our current molecular framework,
suggesting the need for alternative or complementary approaches to
fully account for all observations. 

In conclusion, while our study provides valuable insights into the
nature of hidden charm pentaquarks and offers a potential explanation
for their absence in certain experimental settings, it also highlights
the need for further theoretical and experimental work to
comprehensively understand these exotic particles.

\begin{acknowledgments}
S.C. and H.C.K. wish to express their gratitude to T. Mart at
Universitas Indonesia for his hospitality during their visit to
Depok, where part of the present work was conducted.  
The work was supported by the Basic
Science Research Program through the National Research Foundation of
Korea funded by the Korean government (Ministry of Education, Science
and Technology, MEST), Grant-No. 2021R1A2C2093368 and 2018R1A5A1025563
(SC and HChK), and by the PUTI Q1 Grant from University of Indonesia
under contract No. NKB-441/UN2.RST/HKP.05.00/2024.
\end{acknowledgments}

\bibliography{Pc}
\bibliographystyle{apsrev4-2}

\end{document}